\documentclass[journal=jctc,manuscript=article,layout=twocolumn]{achemso}
 
\usepackage[version=3]{mhchem} 
 \SectionNumbersOn
\usepackage{graphicx}
 
\usepackage{bm}
\usepackage{braket}
\usepackage{amsfonts}
\usepackage{amsmath}
\usepackage{amssymb}
\usepackage[varg]{txfonts}
\usepackage[font=footnotesize,labelfont=bf]{caption}
\usepackage[small]{titlesec}
\DeclareMathOperator{\sgn}{sgn}
\usepackage{xcolor}
 
\definecolor{colorbg}{RGB}{139, 233, 253}
\definecolor{colorkeyone}{RGB}{183, 28, 28}
\definecolor{colorkeytwo}{RGB}{25, 118, 210}
\definecolor{colorkeythree}{RGB}{46, 125, 50}
\definecolor{colorkeyfour}{RGB}{197, 96, 0}
\definecolor{colorkeyfive}{RGB}{64, 36, 26}
 
\definecolor{frgreen}{RGB}{0, 165, 0}
 
\usepackage{algorithm2e}
\SetInd{0.1em}{0.8em}

\SetKwFor{ForEach}{\textcolor{colorkeytwo}{foreach}}{\textcolor{colorkeytwo}{do}}{\textcolor{colorkeytwo}{end}}%
\SetKwFor{For}{\textcolor{colorkeyfour}{for}}{\textcolor{colorkeyfour}{do}}{\textcolor{colorkeyfour}{end}}%
\SetKwFor{While}{\textcolor{colorkeythree}{while}}{\textcolor{colorkeythree}{do}}{\textcolor{colorkeythree}{end}}%
\SetKwIF{If}{ElseIf}{Else}{\textcolor{colorkeyone}{if}}
{\textcolor{colorkeyone}{then}}{\textcolor{colorkeyone}{else if}}
{\textcolor{colorkeyone}{else}}{\textcolor{colorkeyone}{end if}}

\SetCommentSty{mycommfont}
 
\usepackage{booktabs}
 
\usepackage{tikz}
\usetikzlibrary{fit, calc}

\usepackage{float}
\author{Ram\'on L. Panad\'es-Barrueta}
\affiliation{Faculty of Chemistry and Food Chemistry, Technische Universit\"at Dresden, 01062 Dresden, Germany}
\email{ramon_lorenzo.panades-barrueta@tu-dresden.de}
\author{Dorothea Golze}
\affiliation{Faculty of Chemistry and Food Chemistry, Technische Universit\"at Dresden, 01062 Dresden, Germany}
\email{dorothea.golze@tu-dresden.de}
 
\title[cdwac]{Accelerating core-level \textit{GW} calculations by combining
  the contour deformation approach with the analytic continuation of \textit{W}.}
 
\abbreviations{GW,HPC,FHI-aims}
\keywords{Core level spectroscopy, Contour deformation}
 
\let\oldmaketitle\maketitle
\let\maketitle\relax
\begin{document}
\linespread{1.1}
\fontsize{10}{12}\selectfont
\twocolumn[
  \begin{@twocolumnfalse}
    \oldmaketitle
    \begin{abstract}
\fontsize{10}{12}\selectfont  
  In recent years, the \(GW\) method has emerged as a reliable tool for computing core-level binding energies. The contour deformation (CD) technique has been established as an efficient, scalable, and numerically stable approach to compute the \(GW\) self-energy for deep core excitations. However, core-level \(GW\) calculations with CD face the challenge of higher scaling with respect to system size $N$ compared to the conventional quartic scaling in valence state algorithms. In this work, we present the CD-WAC method (CD with $W$ Analytic Continuation), which reduces the scaling of CD applied to the inner shells from \(O(N^5)\) to \(O(N^4)\) by employing an analytic continuation of the screened Coulomb interaction \(W\). Our proposed method retains the numerical accuracy of CD for the computationally challenging deep core case, yielding mean absolute errors $<5~$~meV for well-established benchmark sets, such as CORE65, for single-shot $GW$ calculations. More extensive testing for different \(GW\) flavors prove the reliability of the method. We have confirmed the theoretical scaling by performing scaling experiments on large acene chains and amorphous carbon clusters, achieving speedups of up to 10x for structures of only 116 atoms. This improvement in computational efficiency paves the way for more accurate and efficient core-level \(GW\) calculations on larger and more complex systems.
    \end{abstract}
  \end{@twocolumnfalse}
  ]
 
 
\section{Introduction}\label{sec:intro}
 
Core level spectroscopy, particularly X-ray Photoelectron Spectroscopy (XPS), is a powerful analytical technique, which probes the binding energies of the core electrons. XPS
provides valuable insights into the elemental composition of the material or molecule and into the chemical environment of the core-excited atom~\cite{siegbahnESCAAppliedFree1969,bagusInterpretationXPSSpectra2013}. However, the interpretation of XPS spectra is often challenging due to the lack of reference data, necessitating the use of theoretical calculations~\cite{van2011x}.
 
Methods based on density functional theory (DFT)~\cite{hohenbergInhomogeneousElectronGas1964,
kohnSelfConsistentEquationsIncluding1965}, specifically the Delta self-consistent field ($\Delta$SCF) method~\cite{bagus1965self}, have been widely used for computing XPS spectra~\cite{susi_2015,Susi2018,Bellafont2016,Bellafont2016b,vines2018prediction,aarva_2019,aarva_2019b,kahk2019,Hait2020,besley2021wiley,kahk2021,Klein2021,Kahk2022,Kahk2023}. While DFT is computationally efficient, it is a ground-state theory and does not provide systematic access to electronic excitations. Problems of  $\Delta$SCF related to the inclusion of periodicity, constraining the core hole and self-interaction errors have been discussed elsewhere~\cite{Pinheiro2015,golze2018core,michelitsch2019,golze2022accurate,Hall2021}. 
 
Rigorous theoretical frameworks for exciting electrons are provided by response methods, where electron propagators are applied to transform the ground to an excited state. In the realm of wave-function-based methods, this includes equation-of-motion coupled cluster (EOM-CC)\cite{Stanton1993} and the algebraic diagrammatic construction (ADC) scheme~\cite{schirmer1982beyond,kuleff2014ultrafast}. Recently, both EOM-CC\cite{liuBenchmarkCalculationsKEdge2019,vidal2020,ambroise2021,vila2021} and ADC\cite{norman2018simulating,ambroise2021} were successfully applied to deep core excitations. An alternative approach is the \(GW\) family of methods~\cite{hedinNewMethodCalculating1965,aryasetiawan1998gw,reining2018gw,golze2019gw}, which are generally computationally less expensive than wave-function based approaches. ~\cite{jacquemin20150,bintrim2021full} The fundamental object of the \(GW\) approximation is the one-particle Green's
function or electron propagator (\(G\)), whose imaginary part gives access to the intrinsic spectral function~\cite{golze2019gw}. The latter observable can be directly related to the photocurrent measured in photoemission experiments.
 
\(GW\) has become the method of choice for the calculation of direct and indirect photoemission spectra of solids\cite{golze2019gw,rasmussenFullyAutomatedGW2021} and molecules.\cite{van2015gw,golze2019gw,stukeAtomicStructuresOrbital2020,brunevalGWMiracleManyBody2021}. However, the calculation of deep core-level binding energies ($>100$~eV) is a very recent trend in the \(GW\) community.\cite{aoki2018accurate,van2018assessing,golze2018core,voora2019effective,golze2020accurate,keller2020relativistic,duchemin2020robust,zhu2021all,mejia2021scalable,mejia-rodriguezBasisSetSelection2022,golze2022accurate,li2021renormalized,li2022benchmark,galleni2022modeling,mukatayev2023xps} Lately, \(GW\) has also been employed to compute $K$-edge transition energies measured in X-ray absorption spectroscopy (XAS) by extension to the Bethe-Salpeter equation (BSE@$GW$).\cite{yaoAllElectronBSEGW2022} 
The recent interest in \(GW\) for XPS and XAS computations is driven by its availability in all-electron localized basis set codes, which is a development of the last decade.\cite{golze2019gw}
 
\(GW\) calculations of core levels are more demanding than their valence counterparts. While an all-electron treatment is the basic requirement for core-level \(GW\), the following measures must be taken to obtain reliable results and quantitative agreement with experiment: i) We found that highly exact frequency integration techniques for the calculation of the self-energy are necessary, such as a fully analytic treatment or the contour deformation approach.\cite{golze2018core} Unlike for valence states, the self-energy has complicated features with many poles. Popular valence-state algorithms using an analytic continuation (AC) of the self-energy fail here drastically. ii) Standard \(G_0W_0\) approach using generalized gradient approximations (GGAs) as starting point suffer from an extreme, erroneous transfer of spectral weight to the satellite spectrum.\cite{golze2020accurate} A distinct quasiparticle (QP) solution is \textit{not} obtained at the \(G_0W_0\)@GGA level. We showed that the inclusion of partial eigenvalue-selfconsistency in $G$ (ev$GW_0$)  restores the QP peak. We proposed as computationally cheaper alternatives to ev$GW_0$ either \(G_0W_0\) starting from a hybrid functional with 45\% of exact exchange \cite{golze2020accurate,li2022benchmark} or a so-called Hedin shift in $G$,\cite{li2022benchmark} which can be considered a simplified version of ev$GW_0$. iii) Relativistic effects have also been found to play a major role in the correct description of core-level excitations~\cite{keller2020relativistic}. All three points combined, $GW$ yields excellent agreement with experiment, with errors $<0.2$~eV for absolute and relative binding energies of molecules, respectively.\cite{li2022benchmark}  Moreover, we showed for disordered carbon-based materials that we can resolve spectral features within 0.1 eV of the reference experimental spectra when including the $GW$ correction in the XPS predictions.\cite{golze2022accurate}
 
The numerical requirements described in i) increase the computational demands compared to valence calculations. A fully analytic evaluation of the self-energy 
scales $O(N^6)$ with respect to system size $N$.\cite{van2013gw} The  complexity of core-level $GW$ calculations with CD, denoted as CD-$GW$ in the following, is with $O(N^5)$ lower and the reason we preferred CD over a fully analytic approach previously\cite{golze2018core}. However, this is still higher than in canonical $GW$ implementations with $O(N^4)$ scaling.~\cite{blase2011first,ren2012resolution,van2013gw,deslippe2012berkeleygw,gonze2009abinit,klimevs2014predictive,huser2013quasiparticle,wilhelmGWGaussianPlane2016,wilhelm2017periodic,zhu2021all,ren2021}  Currently, the $O(N^5)$ scaling of CD-$GW$ limits its application to systems of $\approx$ 100–120 atoms. Scaling reduction is thus inevitable in order to address larger system sizes.  
 
The development of low-scaling formulations of $GW$ is an active field of research. The most popular low-scaling approach
is the space-time method~\cite{rojas1995space} which yields cubic scaling algorithms and has gained widespread adoption, as evidenced by a variety of implementations in, e.g., a plane-wave/projector-augmented-wave $GW$ code \cite{liu2016cubic} or with localized basis sets using Gaussian \cite{wilhelm2018toward,wilhelm2021low,duchemin2021cubic} or Slater-type orbitals~\cite{foerster2020low,foerster2021loworder,foerster2023twocomponent}. However, the space-time method relies on a formulation in imaginary frequency and time, whereas we have a dependency on real frequencies in the CD approach. A direct application of the space-time method is thus not straightforward.
 
One possible way of reducing the CD scaling is recasting expensive
tensor contractions into a linear system of equations that can be
solved using, for example, Krylov subspace methods~\cite{wright2006numerical}. This approach is used in Ref.~\citenum{mejia2021scalable}, which uses the minimal residual method (MINRES) algorithm for the aforementioned purpose, effectively achieving a scaling reduction to \(O(N^4)\). Despite its potential, this method comes with two significant drawbacks. Firstly, the solver often requires a substantial number of iterations to attain a desirable level of accuracy. Secondly, due to the intricate structure of the self-energy typically observed in core-level states~\cite{golze2018core}, the method is sensitive to finding solutions which might have a low spectral weight~\cite{mejia2021scalable}.
 
In this work, we present a more streamlined and elegant approach to the scaling reduction of CD-\(GW\) based on the AC of the screened Coulomb interaction $W$, which we denote as CD-WAC (CD with $W$ analytic continuation). While the AC of the self-energy is very common and implemented in many \(GW\) codes~\cite{gonze2009abinit,ren2012resolution,liu2016cubic,wilhelmGWGaussianPlane2016,wilhelm2017periodic,wilhelm2018toward,wilhelm2021low,foerster2020low,foerster2021loworder}, the computational benefits of an AC of $W$ have hardly been investigated. We are only aware of three cases where the latter was used~\cite{friedrich2019tetrahedron,voora2020molecular,samal2022modeling,duchemin2020robust}. Friedrich \textit{et al.}\cite{friedrich2019tetrahedron} introduced the idea in the context of the \(T\)-matrix approach~\cite{springer1998first}. Duchemin and Blase~\cite{duchemin2020robust} explored the CD-WAC idea for valence states, but provided only a preliminary case study for the O1s excitation of a single \ce{H2O} molecule. Voora and collaborators~\cite{voora2020molecular,samal2022modeling} used a similar idea in a generalized Kohn–Sham approach to random phase approximation (RPA) calculations. Here we advance the CD-WAC approach for deep core-level calculations and show that we can elegantly use it to reduce the scaling of CD-$GW$. 
 
The paper is structured as follows: in
Section~\ref{sec:theory}, we provide an overview of the \(GW\) theory,
with strong emphasis on the CD formalism. We then introduce the CD-WAC approach and discuss the choice of the algorithm used for the
AC of \(W\). The Section~\ref{sec:imp} offers
details of the implementation, including a detailed algorithm
description. The computational details are presented in
Section~\ref{sec:comp}. We then show the results of the accuracy and
computational performance benchmarks in section~\ref{sec:res}. Finally, we provide our conclusions and future perspectives of this work in Section~\ref{sec:conc}.
 
\section{Theory}\label{sec:theory}
 
\subsection{The \boldsymbol{$G_0W_0$} approximation}\label{sss:g0w0}
 
The \(GW\) approximation to many-body perturbation theory
(MBPT) is derived from the  Hedin's equations~\cite{hedin1965new,golze2019gw} by omitting the vertex corrections. The most important object in \(GW\) is the self-energy, which accounts for exchange and correlation contributions beyond the Hartree-Fock approximation and is given by
{\footnotesize
  \begin{equation}
    \label{eq:selfew}
    \Sigma(\mathbf{r}_1,\mathbf{r}_2,\omega) =
    \frac{i}{2\pi} \int G(\mathbf{r}_1,\mathbf{r}_2,\omega+\omega')
    W(\mathbf{r}_1,\mathbf{r}_2,\omega')e^{i\omega'\eta} d\omega'
\end{equation}}where \(G\) and \(W\) are the Green's function and
screened Coulomb interaction respectively.
 
The lowest rung in the hierarchy of \(GW\) approximations is the
so-called \(G_0W_0\) approach.  The approach perturbatively improves the self-energy corresponding to a mean-field Green's function \(G_0\) by
performing a single iteration of the simplified Hedin's
equations~\cite{golze2019gw}. A common choice to build \(G_0\) is to
use the eigendecomposition of an initial DFT Hamiltonian. In the Lehmann representation~\cite{gross1986many,schirmer2018many} this can be done as follows:
\begin{equation}
  \label{eq:gnot}
  G_0^{\sigma}(\mathbf{r}_1, \mathbf{r}_2, \omega) = \sum_m
  \frac{\Psi_{m\sigma}(\mathbf{r}_1)\Psi^{*}_{m\sigma}(\mathbf{r}_2)}
  {\omega - \epsilon_{m\sigma}- i\eta \sgn(\epsilon_F-\epsilon_{m\sigma})}
\end{equation}
where \(\epsilon_{m\sigma}\)  and \(\Psi_{m\sigma}\) are the Kohn–Sham (KS)
eigenvalues and eigenvectors for the spin channel \(\sigma\), and \(\epsilon_F\) is the
Fermi level. The non-interacting screened Coulomb
interaction \(W_0\) is computed in the RPA:
\begin{equation}
  \label{eq:w0}
  W_0(\mathbf{r}_1, \mathbf{r}_2, \omega) = \int 
  \varepsilon^{-1}(\mathbf{r}_1, \mathbf{r}_3, \omega)v(\mathbf{r}_3,
  \mathbf{r}_2) d\mathbf{r}_3
\end{equation}
where \(v\) is the bare Coulomb interaction, and
\(\varepsilon^{-1}\) is the inverse of the dielectric function, which
can be computed as:
\begin{equation}
  \label{eq:dielec}
  \varepsilon(\mathbf{r}_1, \mathbf{r}_2, \omega) =
  \delta(\mathbf{r}_1, \mathbf{r}_2) - \int v(\mathbf{r}_1,
  \mathbf{r}_3)
  \chi_0(\mathbf{r}_3, \mathbf{r}_2, \omega)d\mathbf{r}_3
\end{equation}
 
In the previous equation we have introduced the irreducible
polarizability $\chi_0$, which is in practice computed using the Adler-Wisser
expression~\cite{adler1962quantum,wiser1963dielectric}:
{\footnotesize
  \begin{equation}
    \label{eq:aw}
    \begin{alignedat}{2}
      \chi_0(\mathbf{r}_1, &\mathbf{r}_2, \omega) =\\
                           &\sum_{\sigma}\sum_{i}^{\text{occ}}\sum_{a}^{\text{virt}}
                           &&\Biggl\{ \frac{\Psi_{i\sigma}^{*}(\mathbf{r}_1)\Psi_{a\sigma}(\mathbf{r}_1)
                              \Psi_{a\sigma}^{*}(\mathbf{r}_2)\Psi_{i\sigma}(\mathbf{r}_2)}
                              {\omega-(\epsilon_{a\sigma}-\epsilon_{i\sigma})+i\eta}\\
                           & &&-\frac{\Psi_{i\sigma}(\mathbf{r}_1)\Psi_{a\sigma}^{*}(\mathbf{r}_1)
                                \Psi_{a\sigma}(\mathbf{r}_2)\Psi_{i\sigma}^{*}(\mathbf{r}_2)}
                                {\omega+(\epsilon_{a\sigma}-\epsilon_{i\sigma})-i\eta}\Biggr\}
    \end{alignedat}
  \end{equation}}where the indexes \(i, a\) run over occupied and virtual states, 
respectively.
 
For computational reasons as well as for better physical
interpretation~\cite{golze2019gw} the self energy is usually split
into a correlation (\(c\)) and exchange (\(x\)) contribution
\(\Sigma^{\sigma}=\Sigma^{c,\sigma}+\Sigma^{x,\sigma}\). The latter can be expressed in terms of
the bare Coulomb interaction:
{\footnotesize
  \begin{equation}
    \label{eq:exs}
      \Sigma^{x,\sigma}(\mathbf{r}_1, \mathbf{r}_2, \omega) =
     -\sum_i^{occ} \Psi_{i\sigma}(\mathbf{r}_1) \Psi^{*}_{i\sigma}(\mathbf{r}_2)  v(\mathbf{r}_1,\mathbf{r}_2)
  \end{equation}}
The expression for \(\Sigma^{c,\sigma}\) is analogous to that of
Equation~\eqref{eq:selfew}, but using only the correlation part $W^c_0$ of the screened Coulomb
interaction:
\begin{equation}
  \label{eq:scnew}
  W^c_0(\mathbf{r}_1, \mathbf{r}_2, \omega) = W_0(\mathbf{r}_1,
  \mathbf{r}_2, \omega) - v(\mathbf{r}_1, \mathbf{r}_2)
\end{equation}
 
The $G_0W_0$ QP energy $\epsilon^{G_0W_0}_{n\sigma}$ for state $n$ can be computed as first order corrections of the KS eigenvalue $\epsilon_n$ by solving the following fixed-point equation:
\begin{equation}
  \label{eq:qp_eq}
  \epsilon^{G_0W_0}_{n\sigma}= \epsilon_{n\sigma} + \Re\Sigma_n^{\sigma}(\epsilon^{G_0W_0}_{n\sigma}) - v_n^{xc,\sigma}
\end{equation}
where $v^{xc,\sigma}$ is the exchange-correlation potential from KS-DFT and where
\begin{equation}
  \label{eq:vxc_se}
  \begin{aligned}
    v_{n}^{xc,\sigma} &= \int
                        \Psi_{n\sigma}^{*}(\mathbf{r}_1)
                        v^{xc,\sigma}(\mathbf{r}_1)
                        \Psi_{n\sigma}(\mathbf{r}_1)
                        d\mathbf{r}_1\\
    \Sigma_n^{\sigma}(\omega) &= \int
                                \Psi_{n\sigma}^{*}(\mathbf{r}_1)
                                \Sigma^{\sigma}(\mathbf{r}_1,\mathbf{r}_2,\omega)
                                \Psi_{n\sigma}(\mathbf{r}_2)
                                d\mathbf{r}_1d\mathbf{r}_2
  \end{aligned}
\end{equation}
In the following, we will omit the explicit spin dependency in order
to declutter the notation.
 
One of the biggest challenges in \(GW\) calculations is the computation of \(\Sigma^{c}(\mathbf{r}_1,\mathbf{r}_2,\omega)\). The integral in Equation~\eqref{eq:selfew} can be solved analytically (for the correlation part) by computing
\begin{equation}
 \Sigma_n^c(\omega) = \sum_m \sum_{s} \frac{\Braket{\Psi_n\Psi_m|P_s|\Psi_m\Psi_n}}{\omega -\epsilon_m + 
(\Omega_s-i\eta)\sgn(\epsilon_{\mathrm{F}} -\epsilon_m)}
    \label{eq:sigma_pole}
\end{equation}
where $\Omega_s$ are charge neutral excitations and $P_s$ transition amplitudes. This procedure is in principle the most exact way to compute $\Sigma_n^c$, but gives rise to an algorithm with \(O(N^6)\) complexity~\cite{van2013gw,bintrim2021full}. However, there is a set of approximate and exact alternatives available to solve Equation~\eqref{eq:selfew}~\cite{golze2019gw} that yield the \(O(N^4)\) scaling of the canonical \(G_0W_0\) algorithm. 
 
Leaving the plasmon-pole models~\cite{hybertsen1986electron} aside, there are several full frequency integration techniques available. A popular approach is to compute $\Sigma_n^c$ on the imaginary axis, where the self-energy integral is smooth and easy to evaluate, and then analytically continue to the real axis. The AC is usually performed by fitting $\Sigma_n^c(i\omega)$ to a multipole model~\cite{rojas1995space} or by interpolation using a Padé
approximant~\cite{van2015gw,wilhelm2018toward}. The AC methods produce accurate results for valence states,~\cite{van2015gw} but fail for core levels, due to the complicated pole structure of the self-energy matrix elements in the deep core region.~\cite{golze2018core} The CD technique is another full frequency approach, which is suitable for core states and which is described in detail in Section~\ref{sec:cd}.

\subsection{Resolution of the identity}
 
The computation of the self-energy matrix elements \(\Sigma_n^{c}\) requires the quadrature of the electron repulsion integrals (ERIs) that occur in Equation~\eqref{eq:aw}. In this work, these operations are accelerated using the resolution of the
identity approximation with the Coulomb metric~\cite{vahtras1993integral,ren2012resolution} (RI-V). The molecular orbitals (MOs) are expanded in localized atom-centered orbitals \(\{\varphi_{\mu}\}\):
\begin{equation}
  \label{eq:monao}
  \Psi_n(\mathbf{r}_i) = \sum_{\mu}C_{\mu n}\varphi_{\mu}(\mathbf{r}_i)
\end{equation}
where $C_{\mu n}$ are the MO coefficients obtained from the proceeding KS-DFT calculation. In RI we represent the products of MOs in terms of an auxiliary basis set \(\{\phi_P\}\):
\begin{equation}
  \label{eq:rione}
  \Psi_{n}(\mathbf{r}_i)\Psi_{m}(\mathbf{r}_i)=\sum_PA^{nm}_P\phi_P(\mathbf{r}_i)
\end{equation}
The ERIs can then be expressed as:
\begin{equation}
  \label{eq:eristwo}
  \begin{aligned}
    (nm|kl) &= \sum_{PQ}A^{nm}_P(P|Q)A^{kl}_Q\\
    (P|Q) &= \int \phi_P(\mathbf{r}_1)v(\mathbf{r}_1,
            \mathbf{r}_2)\phi_Q(\mathbf{r}_2)d\mathbf{r}_1d\mathbf{r}_2
  \end{aligned}
\end{equation}
 
In RI-V, the tensors \(A^{nm}_P\) are obtained by minimizing the \(L_2\) norm of the RI error of the four center integrals~\cite{vahtras1993integral,ren2012resolution}. We can then present the working equations of RI-V\@:
{\footnotesize
  \begin{equation}
    \label{eq:workriv}
    \begin{aligned}
      {(nm|kl)}_{\text{RI-V}} &= \sum_{PQ}(nm|P)(P|Q)^{-1}(Q|kl)\\
                       &= \sum_{R}\Big\{\sum_{P}\sum_{\mu\nu}(\mu\nu|P)C_{\mu n}C_{\nu
                         m}(P|R)^{-\frac{1}{2}}\\
                       &\qquad\quad\times
                         \sum_{Q}\sum_{\lambda\gamma}(R|Q)^{-\frac{1}{2}}C_{\lambda
                         k}C_{\gamma l}(Q|\lambda\gamma)\Big\}\\
                       &= \sum_R\sum_{\mu\nu}M_R^{\mu\nu}C_{\mu n}C_{\nu
                         m}\times
                         \sum_{\lambda\gamma}M_R^{\lambda\gamma}C_{\lambda      k}C_{\gamma l}\\
                       &=\sum_{R}O^{nm}_{R}O^{kl}_{R}
    \end{aligned}
  \end{equation}}
with the three-center quantities
  \begin{equation}
   M_P^{\mu\nu} = \sum_R(\mu\nu|R)(R|P)^{-1/2}
  \end{equation}
and their transformation in the MO basis
\begin{equation}
   O^{nm}_{P}  = \sum_{\mu\nu}M_P^{\mu\nu}C_{\mu n}C_{\nu m}
   \label{eq:3c_OnmP}
\end{equation}
where the Greek letter indexes \(\mu,\nu,\lambda,\gamma\) refer to the atom-centered orbitals of the primary basis. The effective scaling of the RI-V approximation is \(O(N^3)\). The memory requirements and prefactor of the algorithm also get dramatically reduced~\cite{ren2012resolution}, as one only needs to compute three and two center integrals, and only the \(O_{R}^{nm}\) tensors are stored.
 
\subsection{Contour deformation technique}\label{sec:cd}
 
The CD technique~\cite{godby1988self,govoni2015large,golze2018core} has been successfully applied to both valence and core electrons. To name but a few recent applications, consult
Refs.~\citenum{golze2018core,keller2020relativistic,wilhelm2021low,yao2022all,nabok2022bulk,li2022benchmark}. Moreover, we showed for deep core states that the CD self-energies exactly match the fully analytic results from Equation~\eqref{eq:sigma_pole}.\cite{golze2018core} A full derivation of the CD approach in combination with RI-V has been given in our previous work.\cite{golze2018core} We summarize in the following the basic idea and the final expressions.
 
\begin{figure}[t]
    \centering
    \includegraphics[scale=.6]{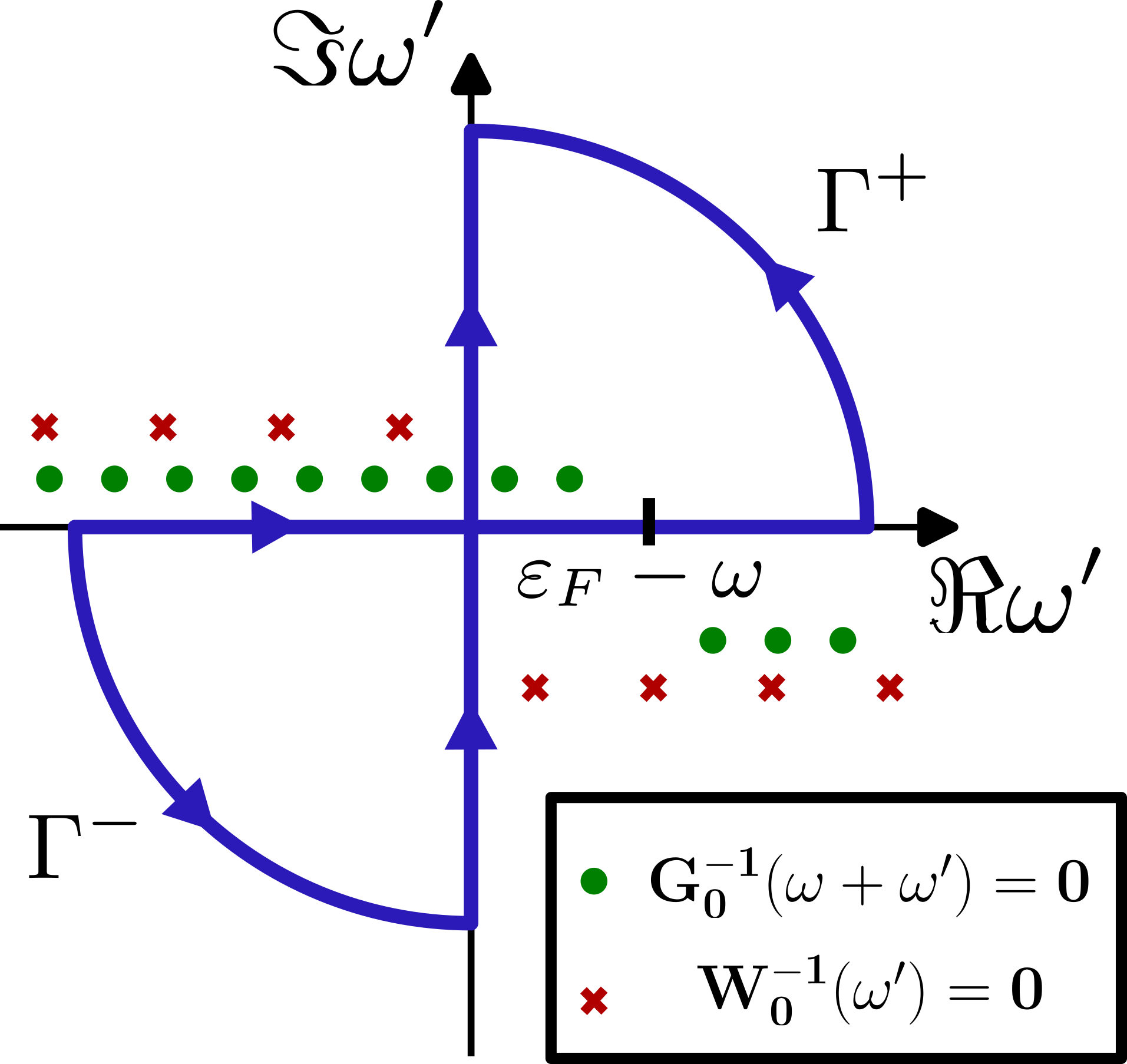}
    \caption{Integration path employed for the complex quadrature of the self-energy matrix elements \(\Sigma_n(\omega)\) in the CD method. By design, only some poles of \(G_0(\omega+\omega')\) (green dots) lay inside the bounded region, and none of the \(W_0(\omega')\) poles (red crosses).}
    \label{fig:cd_path}
\end{figure}

The central idea of the method is to compute the self-energy on the real frequency axis using an integration contour that minimizes the amount of residues inside the bounded area. In particular, all poles of \(W\) lay outside of it and only some poles of \(G\) enter the first or third quadrants. The chosen contour is represented in Figure~\ref{fig:cd_path}.
An application of Cauchy's residue theorem and Jordan's lemma for the \(\Gamma^{+}\) and \(\Gamma^{-}\) arcs readily gives:
\begin{equation}
\Sigma^{c}(\mathbf{r}_1,\mathbf{r}_2,\omega) =
                  -I^c(\mathbf{r}_1,\mathbf{r}_2,\omega)
                  +R^c(\mathbf{r}_1,\mathbf{r}_2,\omega)
\label{eq:cdmain}
\end{equation}
with the definitions:
\begin{equation}
\label{eq:cdmain_defs_a}
    I^c =-\frac{1}{2\pi}\int_{-\infty}^{+\infty}d\omega'
    G_{0}(\mathbf{r}_1,\mathbf{r}_2,\omega+i\omega')
    W_{0}^c(\mathbf{r}_1,\mathbf{r}_2,i\omega')\\
\end{equation}
\begin{equation}
\label{eq:cdmain_defs_b}
    \begin{aligned}
        R^c &= - \sum_j^{\text{occ}}
        \Psi_{j}(\mathbf{r}_1)\Psi_{j}^{*}(\mathbf{r}_2)
        W_{0}^c(\mathbf{r}_1,\mathbf{r}_2,
        \epsilon_{j}-\omega+i\eta)\theta(\epsilon_j-\omega)\\
        & \quad + \sum_a^{\text{virt}} \Psi_a(\mathbf{r}_1)\Psi_a^{*}(\mathbf{r}_2)
        W_{0}^c(\mathbf{r}_1,\mathbf{r}_2,\epsilon_a-\omega-i\eta)\theta(\omega-\epsilon_a)
    \end{aligned}
\end{equation}
  
Using the RI-V approximation, we can compute the corresponding matrix elements \(I_n^c\) and \(R_n^c\) as follows:
\begin{align}
    I_n^c(\omega) &= \frac{1}{2\pi}\sum_m\int
                    \frac{W_{nm}^c(i\omega')d\omega'}
                    {\omega+i\omega'-\epsilon_m -i\eta\,\text{sgn}(\epsilon_F-\epsilon_m)}\label{eq:In_term}\\
    R_n^c(\omega) &= \sum_mf_m W_{nm}^c(|\epsilon_m-\omega|+i\eta)\label{eq:Rn_term}
  \end{align}
where the factors \(f_m\) are defined in terms of Heaviside step functions:
\begin{equation}
  \label{eq:ffact}
  f_m= -\theta(\epsilon_F-\epsilon_m)\theta(\epsilon_m-\omega)+
  \theta(\epsilon_m-\epsilon_F)\theta(\omega-\epsilon_m)
\end{equation}
and the matrix elements \(W_{nm}^c\) are computed for an arbitrary complex argument \(\omega\) as:
 
\begin{equation}
  \label{eq:wriv}
    W_{nm}^c(\omega) = \sum_{PQ}O_{P}^{nm}\Lambda_{PQ}(\omega)O_{Q}^{mn}
\end{equation}
with
\begin{equation}
    \bm{\Lambda} =[\bm{I}-\bm{\Pi}(\omega)]^{-1}-\bm{I}
    \label{eq:bigLambda}
\end{equation}
In Equation~\eqref{eq:bigLambda}, we have introduced the polarizability matrices:
\begin{equation}
  \label{eq:piriv}
  \begin{aligned}
    \Pi_{PQ}(\omega) &= \sum_{ia}O_P^{ia}\Bigl[
                       \frac{1}{\omega-(\epsilon_a-\epsilon_i)+i\eta}\\
                     &-
                       \frac{1}{\omega+(\epsilon_a-\epsilon_i)-i\eta}
                       \Bigr]O_Q^{ia}
  \end{aligned}
\end{equation}
Finally, the self-energy matrix elements needed to compute the
QP energies in Equation~\eqref{eq:qp_eq} are:
\begin{equation}
  \label{eq:cdselfe}
  \Sigma_n(\omega) = \Sigma_n^{x}+R_n^c(\omega)-I_n^c(\omega)
\end{equation}
 
Compared to the AC of the self-energy, the complexity of the CD algorithm is the same with a bit larger prefactor, whereas for core states the scaling with respect to system size increases by an order of magnitude. The culprit is the residue term $R_n$, which becomes the computational bottleneck in core-level calculations, see Section~\ref{sec:performance} for a detailed time complexity analysis.
 
\subsection{CD-WAC: Analytic continuation of W}\label{sec:acw}
 
 AC techniques are ubiquitous in \(GW\), but only for the treatment of the self-energy matrix elements. However, they can be used for any meromorphic function, such as the screened Coulomb interaction matrices. Motivated by previous works from
Friedrich~\cite{friedrich2019tetrahedron} and Duchemin \emph{et al.}~\cite{duchemin2020robust}, the idea of the CD-WAC method is to perform an AC of the \(W_{nm}^{c}(|\epsilon_m-\omega| + i\eta)\) matrices in the \(R_n^{c}\) term in Equation~\eqref{eq:Rn_term}. 
The strategy for performing the AC of $W$ is in principle equivalent to that of the self-energy. We compute $W_{nm}^c$ for a set of imaginary or complex reference frequency points and continue to the real axis. 
 
\(W_{nm}^c\) is a quantity that has singularities, i.e. poles at certain frequencies. We use here Padé approximants~\cite{george1975essentials} to derive analytic multipole models for \(W_{nm}^c\). Padé approximants of order $M$ are rational functions of the form:
\begin{equation}
    T^{M}(z)=\frac{A_0 + A_1z + \cdots + A_pz +\cdots + A_{\frac{M - 1}{2}}z^{\frac{M - 1}{2}}}{1 + B_1z + \cdots + B_pz +\cdots +B_{\frac{M}{2}}z^{\frac{M}{2}}}
    \label{eq:pad_old}
\end{equation}
where \(A_p\) and \(B_p\) denote complex coefficients. Equation~\eqref{eq:pad_old} can be represented as a continued fraction,\cite{vidberg1977solving} i.e., an
 expression of the form \(a_1 + 1/(a_2 + 1/(a_3 + 1/(a_p + \cdots)))\) for some other complex coefficients \(a_p\). Using the form of a continued fraction offers computational advantages, as it enables the use of recurrence formulas in place of explicit polynomial fitting~\cite{chui1989multivariate}. Moreover, recurrence formulas are favorable as their computation can be carried out efficiently using dynamic programming techniques, such as memoization.
 
We employ a variant of Thiele's reciprocal differences' algorithm~\cite{milne2000calculus} for the interpolation using the continued fraction. The approximant \(T_{nm}^{M}(z)\) of order $M$ is defined as:
\begin{equation}
W_{nm}^c(z) \approx   T_{nm}^{M}(z) = \cfrac{a_1}{1+\cfrac{a_{2}(z^2-\tilde{z}^2_{1})}
   {\ddots 1 + \cfrac{a_p(z^2-\tilde{z}^2_{p-1})}{1 +
   \cfrac{(z^2-\tilde{z}^2_p)g_{p+1}(z)}{\ddots 1 + a_M(z^2-\tilde{z}^2_{M-1})}
   }}}\\
  \label{eq:pade}
\end{equation}
where $z$ is a complex argument, $a_p$ are complex coefficients which need to be determined, and \(\{\tilde{z}_p\}\) is a set of (possibly complex) frequencies for which we compute the reference $W_{mn}^c$ matrix elements. For the sake of completeness, the equivalence between Equations \eqref{eq:pad_old} and \eqref{eq:pade} is described in detail in Section 2 of the Supporting Information (SI). 
 
For a given set of \(M\) reference frequencies, the following equalities hold:
\begin{equation}
  T_{nm}^M(\omega_i) = W_{nm}^c(\omega_i)  \qquad i = 1,\ldots,M
\end{equation}
The coefficients \(a_i\) can be efficiently computed by recursion
\cite{vidberg1977solving}, 
as follows:
\begin{equation}
  a_i = g_i(\omega_i) \qquad i = 1,\ldots,M
  \label{pre_rec}
\end{equation}
where the functions \(g_p(z)\) are given by:
\begin{equation}
  g_p(\omega_i) =
  \begin{cases}
    W_{nm}^c(\omega_i) & p = 1 \\
    \displaystyle \frac{g_{p-1}(\omega_{p-1})-g_{p-1}(\omega_i)}{(\omega^2_i-\omega^2_{p-1})g_{p-1}(\omega_i)} & p > 1
  \end{cases}
  \label{rec_eq}
\end{equation}
and the index \(p\) runs in the range \(1,\ldots,M\).
 
The argument \(z\) is squared in order to enforce the parity of the screened Coulomb interaction~\cite{duchemin2020robust}, which is an even functions of the frequency \(W(\omega)=W(-\omega)\). Squaring the argument disrupts the appropriate asymptotic behavior of the approximant, which should converge to zero for extremely large frequencies (see Equations~\eqref{eq:wriv} to~\eqref{eq:piriv}). Indeed, in the limit of infinite frequency argument, the fraction \(T_{nm}^{M}(z)\) will only tend to zero for even \(M\), and will yield a finite non-zero constant for odd \(M\). However, we compute QP energies for real-world systems with finite core-level binding energies of several hundred or several thousand electronvolts. In practice, we observe that interpolations derived from specific regions of the complex plane yield highly precise results. Further details can be found in the discussion in Section~\ref{sec:core}.
 
\section{Implementation details}\label{sec:imp}
 
\subsection{Basis sets}\label{secp:imp:bset}
 
The treatment of deep core level requires the usage of basis sets that are able to represent the rapid oscillations occurring within the wave function near the atomic nuclei. For this reason, the CD-WAC method has been implemented in the FHI-aims all-electron software package~\cite{blum2009ab,havu2009efficient}, building on our previous CD implementation.\cite{golze2018core} In FHI-aims, the MOs are represented as linear combinations of numerical atom-centered orbitals (NAOs), which are of the general form~\cite{blum2009ab}:
\begin{equation}
  \label{eq:naos}
  \varphi_{i[lm]}(\mathbf{r}) = \frac{u_i(r)}{r}Y_{lm}(\Omega)
\end{equation}
where \(u_{i}\) is a radial function, and \(Y_{lm}(\Omega)\)
represents the real (\(m>=0\)) and imaginary (\(m<0\)) parts of complex spherical harmonics~\cite{zhang2013numeric}. We remark here that Gaussian type orbitals (GTOs) or Slater type orbitals (STOs) can be also used in FHI-aims, and as can be inferred from Equation~\eqref{eq:naos}, they can be considered a special case of the NAOs.
 
\subsection{Padé approximant construction}\label{secp:imp:ialg}   \label{subsec:padeimp}
 
While the interpolation algorithm used to compute the Padé approximant of \(W\), as outlined in Equations~\eqref{eq:pade}-\eqref{rec_eq}, is computationally efficient, it is susceptible to numerical instability~\cite{graves1980practical}. 
The numerical errors in the QP energies are found to be 100~meV larger on average for core levels than for valence states. This is due to two key reasons: i) the complex pole structure necessitates the use of more interpolation points, and ii) the larger reference frequencies associated with core levels, which are squared in the computation of the \(g_p(\omega_i)\) matrices, may introduce potential precision loss in the calculation. 
 
\begin{figure}[t]
    \centering
    \includegraphics[scale=.9]{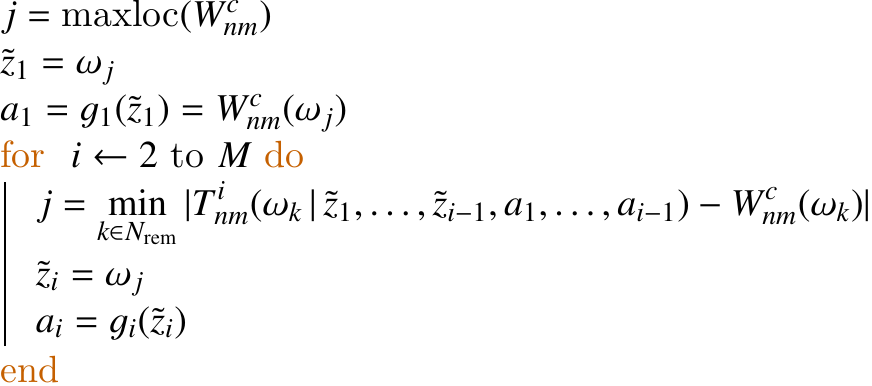}
    \caption{Greedy implementation of the Thiele's reciprocal differences' algorithm. The intrinsic procedure $\texttt{maxloc}$ returns the index of the maximum of the input array. Arguments appearing after the pipe symbol (\(\mid\)) are constant parameters in the function call. See the main text for a discussion of the algorithm steps.}\label{al:inter}
\end{figure}

The numerical instabilities can be mitigated by modifying the sequence in which the reference points are incorporated into the interpolation, as indicated in prior work~\cite{celis2021numerical,celis2023adaptive}. Utilizing this approach, we have successfully stabilized the calculation of the Padé approximant and implemented the method using the greedy algorithm depicted in Figure~\ref{al:inter}.
 
Greedy algorithms make the locally optimal choice at each step, and this becomes apparent in our procedure: giving a number of reference frequencies \(M\), the convergents \(T_{nm}^i\) are constructed by selecting in each iteration the next reference frequency point \(\tilde{z}_i\) from the set of \(N_{\text{rem}}\) remaining points not already included in the convergent \(T_{nm}^{i-1}\). The heuristic criterion that we employed is to select always the \(\omega_k\) that minimizes the \(L_1\) error between the current convergent and the reference matrix value \(W_{nm}^c(\omega_{k})\). In the first iteration step, we always chose the maximum value of \(W_{nm}^c\) as it accounted for better results in the interpolations. We find that our algorithm improves the quality of the QP energies by two orders of magnitude with respect to the standard Thiele's reciprocal differences approach. The results can be further improved by using multiple precision arithmetic, but this implies the use of external libraries that might add technical overhead to the development process, and it is technically more involved than our simpler approach.
 
\subsection{Choice of reference point}\label{secp:imp:rp}
 
CD-WAC requires a set of reference frequencies \(\mathcal{F}_{\text{wac}}\) and the corresponding reference matrix elements $W_{nm}^c$. We have established 
a heuristic procedure to select \(\mathcal{F}_{\text{wac}}\), which is displayed in Figure~\ref{fig:cdwac_regions}.
First, due to the parity of the screened Coulomb interaction function, we only consider points on the first quadrant of the complex plane. Depending on the electronic level studied and the target accuracy, three different regions can be used:
\begin{figure}[t]
  \centering
  \includegraphics[scale=.74]{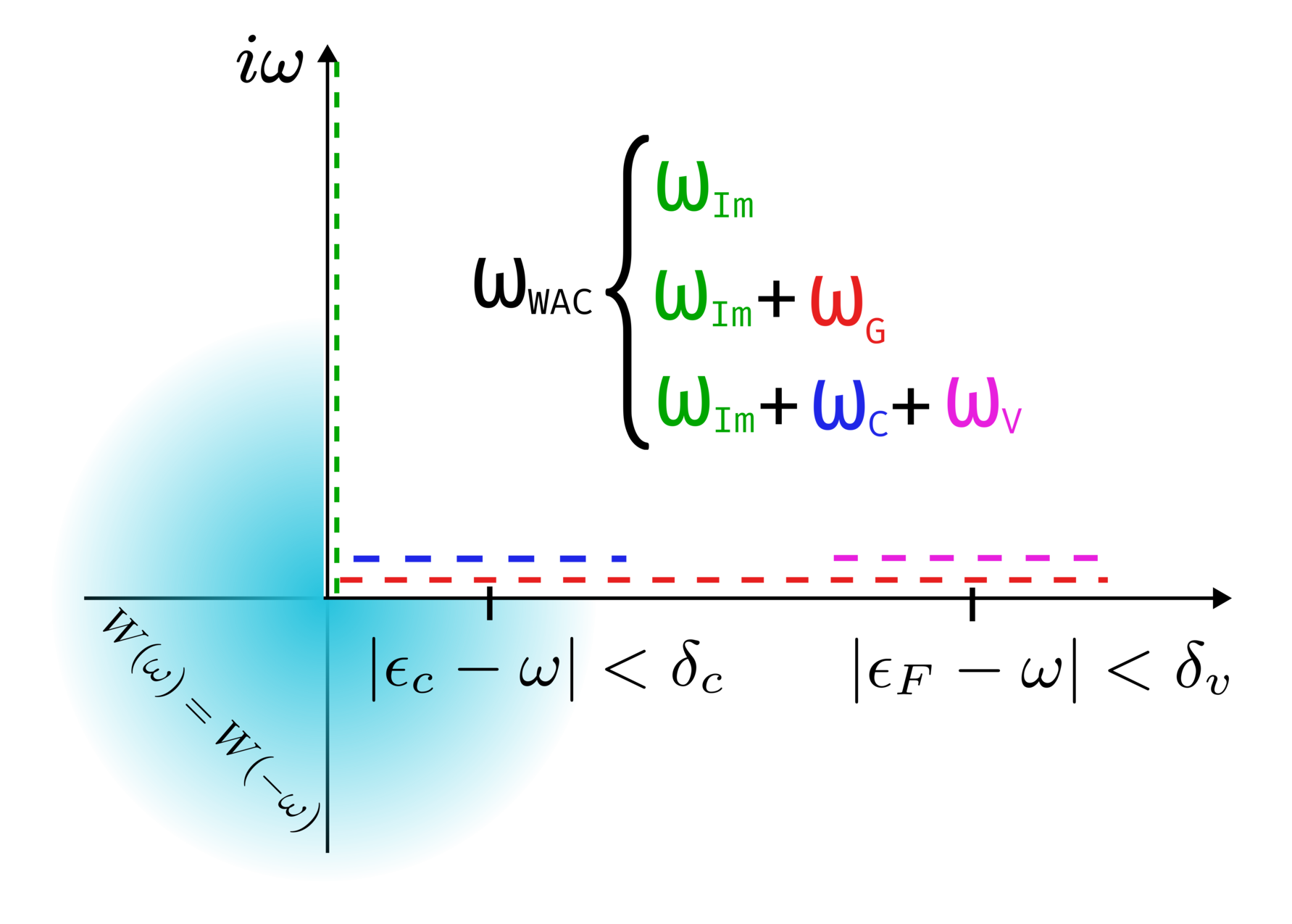}
  \caption{Selection of the reference frequency points for the
    CD-WAC algorithm. The real axis spans the residues
    \(|\epsilon_m-\omega|\), and the imaginary axis encompasses the
 grid used for the numeric integration in \(I_n^c(\omega)\). The
    \(\delta_c\) and \(\delta_v\) parameters control the radii of
    neighborhoods centered at a typical core level residue
    \(|\epsilon_c-\omega|\) and at
    \(|\epsilon_F-\omega|\)
    respectively.}\label{fig:cdwac_regions}
\end{figure}
 
\begin{itemize}
\item \textbf{Valence states}: we reuse the matrix elements \(W_{nm}(i\omega)\) needed for the numerical integration in the \(I^c_n\) term in Equation~\eqref{eq:cdselfe}. These are already computed using the set of imaginary frequencies \(\mathcal{F}_{\text{Im}}=\{{\color{frgreen}{\omega_{\text{Im}}}}\}\). We do not add any extra points. \(\mathcal{F}_{\text{wac}}=\mathcal{F}_{\text{Im}}\)
 
\item \textbf{Core states, generic settings}: we add to  the set of frequencies \(\mathcal{F}_{\text{Im}}\)
additional complex points infinitesimally closed to the real axis  \(\mathcal{F}_{\text{real}}=\{{\color{red}{\omega_{\text{G}}}}\}\), using a grid that spans all the possible residues values up to the Fermi level. The complex part of the additional frequencies is chosen to be a small constant (approx 10$^{-3}$ a.u.), which does not affect the results as long as it is small enough.
\(\mathcal{F}_{\text{wac}}=\mathcal{F}_{\text{Im}}\bigcup \mathcal{F}_{\text{real}}\)
 
\item \textbf{Core states, refined settings}: we add to  the set of frequencies \(\mathcal{F}_{\text{Im}}\)
additional complex points infinitesimally closed to the real axis spanning two different regions \(\mathcal{F}_{\text{real}}=\{{\color{blue}{\omega_{\text{C}}}}\}\bigcup\{{\color{magenta}{\omega_{\text{V}}}}\}\). The first region is a neighborhood centered on a typical core-level residue \(|\epsilon_c-\omega|\). The second one is  a neighborhood centered on  \(|\epsilon_F-\omega|\), representative for a valence-level residue. The radii  of the neighborhoods (\(\delta_c, \delta_v\) respectively) are parameters that can be tuned for improving the interpolation results.
\end{itemize}
 
The purpose of introducing next to generic the refined settings is to minimize the prefactor of the CD-WAC approach and also to increase accuracy by selecting the most relevant reference points, see Section~\ref{sec:calib} for comprehensive convergence tests.  In what follows, we denote the amount of points in the sets
\(\{\omega_C\}\), \(\{\omega_V\}\), and \(\{\omega_{\text{Im}}\}\) as \(N_{\text{core}}\), \(N_{\text{valence}}\), and \(N_{\text{Im}}\) respectively. We also introduce the notation CD-WAC(\(N_{\text{core}}, N_{\text{valence}}, N_{\text{Im}}\)) to denote a CD-WAC calculation carried out with the specified amount of  reference frequencies in the corresponding regions. The order of the Padé approximant will be then given by \(M=N_{\text{core}}+N_{\text{valence}}+N_{\text{Im}}\).
 
\subsection{CD-WAC procedure}\label{secp:imp:wacproc}
 
\begin{figure}
  
    \centering
    \includegraphics{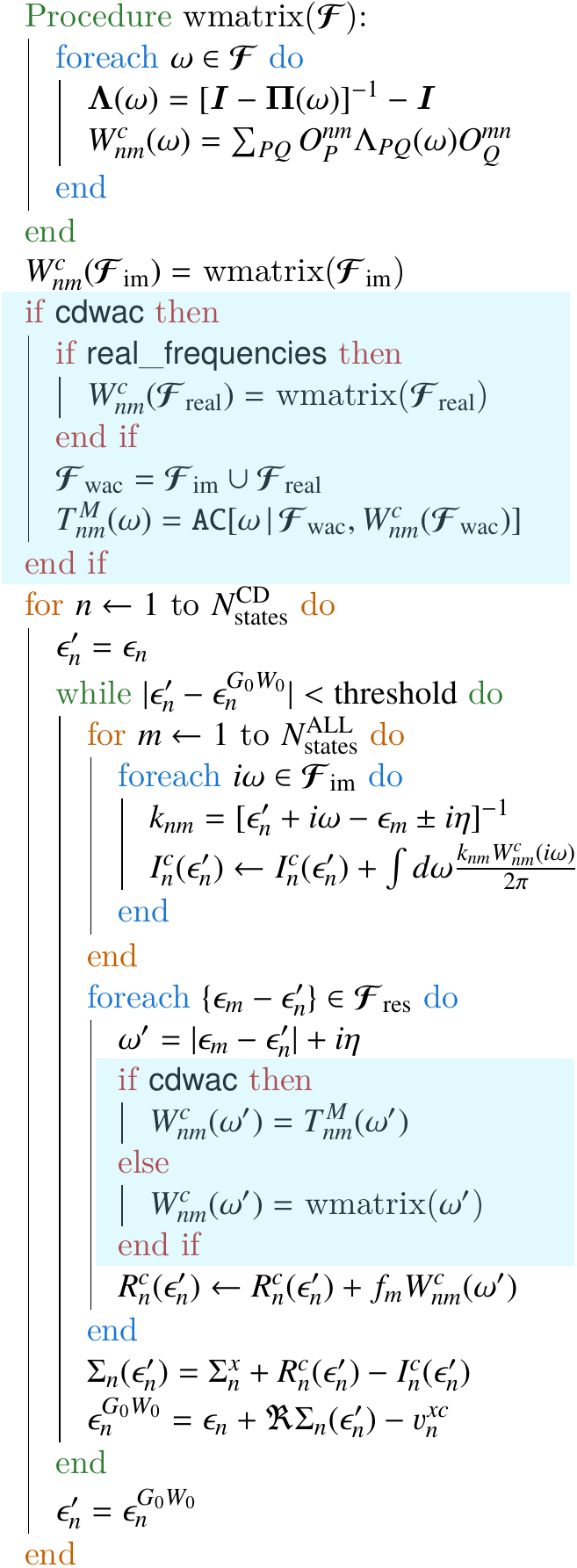}
    \caption{The \(G_0W_0\) algorithm using CD and CD--WAC\@. The cyan
    regions highlight the differences between both approaches. A
    detailed description of the quantities computed in each step can
    be found in the main text.}\label{al:cdwac}
\end{figure}

Figure~\ref{al:cdwac} shows both the CD and CD-WAC algorithm for a \(G_0W_0\) calculation. The differences between the two approaches are highlighted in cyan. We start by defining the \(\mathtt{wmatrix}\) procedure which takes a set \(\bm{\mathcal{F}}\) of frequency points as input and computes the corresponding \(W_{nm}^{c}\) matrix elements. Both methods start then with the computation of \(W_{nm}^{c}(i\omega)\), using a set of imaginary frequencies \(\bm{\mathcal{F}}_{\mathrm{im}}\). The next step is exclusive to CD-WAC: the additional real frequency points are selected, followed by the computation of $W_{nm}^c(\bm{\mathcal{F}}_{\mathrm{real}})$. The AC is performed by the \(\mathtt{AC}\) procedure, which implements the greedy algorithm in Figure~\ref{al:inter}. We use in this procedure the \(M\) frequencies contained in the set \(\bm{\mathcal{F}}_{\mathrm{wac}}\) and the corresponding matrices $W_{nm}^c(\bm{\mathcal{F}}_{\mathrm{wac}})$. Next, given a number of states \(N_{\mathrm{states}}^{\mathrm{CD}}\) to be
treated with CD, the QP equation is solved for each one of them as a fixed point iteration with a certain accuracy threshold. The computation of the \(I_n^c(\epsilon'_n)\) term is carried out in the same fashion for both CD and CD-WAC. We perform a numerical integration and  sum up over $m$ up to the full number of states \(N_{\mathrm{states}}^{\mathrm{ALL}}\). 
 
The second important difference between both approaches arises in
the computation of the \(R_n^c(\epsilon'_n)\) term, also highlighted
in cyan in the algorithm. For the set \(\bm{\mathcal{F}}_{\mathrm{res}}\) of residues specified by Equation~\eqref{eq:ffact}, the CD method computes the matrices \(W_{nm}^c(\omega')\) for the residue \(\omega'\) using the procedure \(\mathtt{wmatrix}\), whereas CD-WAC employs the already constructed approximant \(T_{nm}^M(\omega')\). This step of the algorithm is the computational bottleneck in CD calculations of core excitations. With the introduction of CD-WAC, this step is practically free, albeit at the cost of increasing the prefactor of the procedure before the start of the QP iteration (first cyan box). As a final step, the self-energy matrix elements \(\Sigma_n(\epsilon'_n)\) are computed using \(R_n^c(\epsilon'_n)\) and \(I_n^c(\epsilon'_n)\). Subsequently, they are employed in the update of the QP energies. 
 
Our implementation is designed to be high-performance and capable of handling massive parallelism, making it suited for large scale computations. We leverage distributed parallelism as our primary programming model, maximizing efficiency and scalability. This is achieved by incorporating both established libraries such as ScaLAPACK~\cite{choi1996scalapack} and our own implementation of block cyclic distributions. The latter employs Message Passing Interface~\cite{mpi40} (MPI) for performing matrix multiplication. 
 
\section{Computational details}\label{sec:comp}
 
All $GW$ and underlying DFT calculations are carried out with the FHI-aims program package\cite{blum2009ab}. For the frequency integration of the $GW$ self-energy, we use the CD and CD-WAC\@ techniques. In both cases, the imaginary axis integral (Equation~\eqref{eq:In_term}) is solved by quadrature using a modified Gauss-Legendre grid~\cite{ren2012resolution} of 200 points. 
 
Two benchmark sets are employed to verify the accuracy of CD-WAC with respect to CD, namely CORE65\cite{golze2020accurate} and GW100.\cite{van2015gw} The former includes 65 1s core level excitation energies of 32 organic molecules containing the elements H, C, N, O, and F. The latter reports HOMO and LUMO QP energies from 100 closed-shell molecules covering a wide range of ionization potentials.  The performance of the new implementation was studied using small-medium size organic molecules, in particular we used acene chains~\cite{golze2018core} of the type \ce{(C_{4n+2}H_{2n+4})_{n=1,\ldots,15}}, and a hydrogenated amorphous
carbon material~\cite{golze2022accurate} with the elemental composition \ce{a-C46H70}.
 
The CORE65 benchmark calculations are performed with different \(GW\) flavors. These are i) single-shot  $G_0W_0$ using an optimized starting point, ii) partial and iii) full eigenvalue-self-consistent schemes and iv) a Hedin-shift in the Green's function. We denote the partial eigenvalue-self-consistent scheme as ev\(GW_0\) and the full eigenvalue-self-consistent as ev\(GW\). In ev\(GW_0\), the eigenvalues are only updated in $G$, while in ev\(GW\) the eigenvalues are updated in  $G$ and $W$. The Hedin shift approximates the ev\(GW_0\) scheme by introducing a shift $\Delta\text{H}_n$ in the denominator of $G$. The shift $\Delta\text{H}_n$ is separately determined for each core-level $n$ before starting the iteration of the QP equation (Equation~\eqref{eq:qp_eq}). This approach was comprehensively described in Ref.~\citenum{li2022benchmark} and we refer to it as \(G_{\Delta \text{H}}W_0\) in the following. 
 
We use the Perdew-Burke-Ernzerhof (PBE) functional~\cite{perdew1996generalized} as DFT starting point for the core-level calculations with ev\(GW_0\), ev\(GW\) and \(G_{\Delta \text{H}}W_0\). The core-level $G_0W_0$ calculations are performed on-top of the PBEh($\alpha$) hybrid functional~\cite{atalla2013hybrid} with 45\% of exact exchange ($\alpha=0.45$), which was previously optimized to match the  ev\(GW_0\)@PBE reference and experimental XPS data\cite{golze2020accurate}. For the valence-level excitations, we use $G_0W_0@$PBE as in the original GW100 benchmark paper.\cite{van2015gw} 
 
Relativistic effects were included for all core-level calculations by adding a corrective term derived in Ref.~\citenum{keller2020relativistic} to the non-relativistic $GW$ QP energies. The CORE65 and amorphous
carbon calculations use the Dunning basis sets
family~\cite{dunning1989gaussian,wilson1996gaussian} cc-pVnZ with
\(n\in[\text{T}, \text{Q}]\).  In the case of GW100 and the acene
chains, we employ the def2-QZVP basis set~\cite{weigend2005balanced}. The cc-pVnZ and def2-QZVP are contracted Gaussian basis sets, which can be considered as a special case of an NAO and which are then treated numerically in FHI-aims. The self-energy and screened Coulomb
interaction calculations for the \ce{H2O} molecule were carried out using both the cc-pVQZ and NAOs of the Tier 1 of quality~\cite{blum2009ab} basis sets.  In all cases, RI-V~\cite{vahtras1993integral} was used. In FHI-aims, the auxiliary basis sets for RI-V are constructed on-the-fly by generating on-site products of primary basis functions, which are then orthonormalized for each atom with a Gram-Schmidt procedure.~\cite{ren2012resolution} The input and output files of all the FHI-aims calculations are available in the NOMAD database~\cite{nomaddata}. 
 
\section{Results}\label{sec:res}
 
In the following section, we will analyze our implementation of the CD-WAC method. We will benchmark its accuracy compare to the parent method (CD), and demonstrate that the theoretical speedup is achieved in practice.
 
\subsection{CD-WAC sensible defaults}\label{sec:calib}
 
The first step in our CD-WAC study is the selection of the sensible defaults of the method. That is, the settings that should provide accurate results up to a given threshold and that minimize the number of reference matrix elements $W_{nm}^c$ that must be computed. As decision criterion, we compare the QP energies obtained with CD-WAC to the CD reference.
 
We start with the defaults for the valence states by assessing the GW100 benchmark set using only the imaginary frequency grid
\(\bm{\mathcal{F}}_{\mathrm{im}}\). We find that the mean absolute errors (MAE) for both the HOMO and LUMO states with respect to CD are in the range of \(\sim 10^{-4}\) eV, which is further analyzed in Section~\ref{sec:gw100}. Including only \(\bm{\mathcal{F}}_{\mathrm{im}}\) in the AC of \(W\) seems to be sufficient for frontier orbitals, which is in agreement with the results of Duchemin and Blase~\cite{duchemin2020robust}.
 
For deep core level states, we found that the CD-WAC calculations are more complicated than for valence states, which is also in agreement with the preliminary studies by Duchemin and Blase~\cite{duchemin2020robust}. They studied the 1s core state of a single \ce{H2O} molecule and proposed to complete the imaginary frequency grid \(\bm{\mathcal{F}}_{\mathrm{im}}\) with additional complex frequencies very close to the real axis. We employ a similar, but more refined strategy by using
the heuristic procedure described in Section~\ref{sec:acw}.  
 
We distinguish between generic and refined settings. In the generic brute force approach, the points \(\bm{\mathcal{F}}_{\mathrm{real}}\) are  equidistantly spread, yielding the grid  \(\omega_{G}\). Using a grid size of 100 points for \(\omega_{G}\), we obtain an MAE of 5~meV with respect to CD for the CORE65 benchmarks set. The MAE can be systematically decreased by increasing the size of \(\omega_{G}\), see Figure S1.  While the generic settings yield accurate results, we aim to reduce the total number of points \(\bm{\mathcal{F}}_{\mathrm{real}}\) and the number of outliers by using the refined setting with the two different grids \(\{\omega_{C}\}\) and \(\{\omega_{V}\}\). 
 
\begin{figure}[t]
  \centering
  \includegraphics[scale=.36]{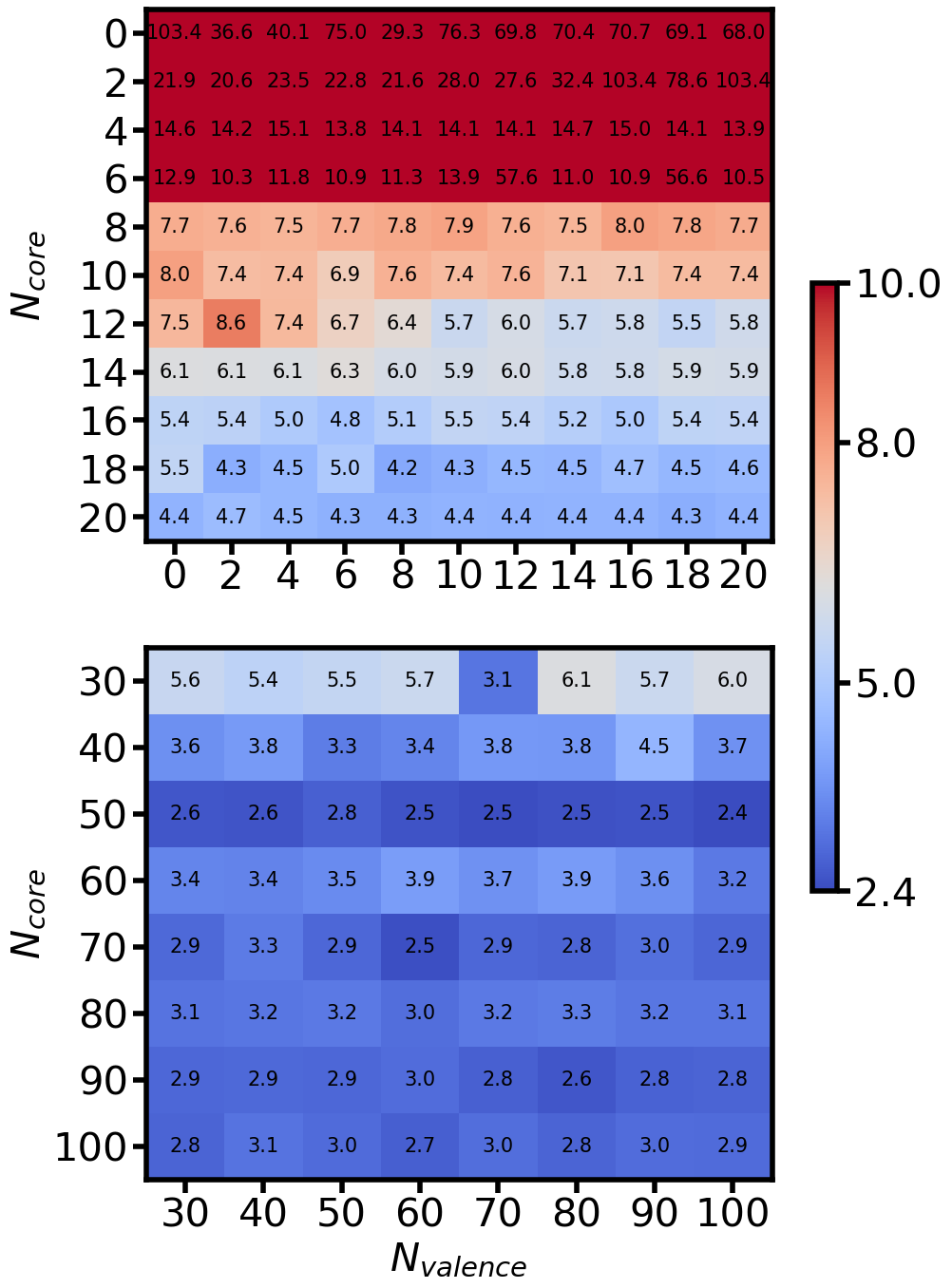}
  \caption{MAEs in meV of CD-WAC with respect to CD for the CORE65 benchmark set as a
    function of the number of additional frequencies \(\bm{\mathcal{F}}_{\mathrm{real}} = \{\omega_{C}\}\bigcup\{\omega_{V}\}\). \(N_{\text{core}}\) and \(N_{\text{valence}}\) are the number of points in \(\{\omega_{C}\}\) and \(\{\omega_{V}\}\), respectively. The 1s QP energies are computed at the $G_0W_0@$PBEh(\(\alpha=0.45\)) level using the cc-pVTZ basis set. 
    \label{fig:hmap}}
\end{figure}

The optimal amount of points \(N_{\text{core}}\) and \(N_{\text{valence}}\) in the sets  \(\{\omega_{C}\}\) and \(\{\omega_{V}\}\), respectively, is determined by computing the MAE with respect to CD for the CORE65 benchmark set at the $G_0W_0@$PBEh(\(\alpha=0.45\)) level using the cc-pVTZ basis set. Keeping a constant size of the imaginary frequency grid \(N_{\text{Im}}=200\),
we explored numerous combinations of points from both
regions. The corresponding MAEs are included in the heat maps in Figure~\ref{fig:hmap}. We studied the progressive increase of
\(N_{\text{core}}\) and \(N_{\text{valence}}\) using fine grids up to
20 points (upper heat map in Figure~\ref{fig:hmap}), and using coarse grids from 20 to 100 points (lower heat map in Figure~\ref{fig:hmap}). 
 
We can see that the MAEs in the coarse grid heat map are
quite stable, with no major improvement being observed when going beyond
30 points for any of the two regions. Larger changes are observed for less than 20 points in both \(\{\omega_{C}\}\) and \(\{\omega_{V}\}\). The heat map with the fine grid displays a decrease of the MAE from  0.103~eV to 0.004~eV with increasing grid size. For most systems, using
\(N_{\text{core}}\sim20\) offers the best results. The dependency on
\(N_{\text{valence}}\) seems to be weaker, although non
negligible, as it systematically improves the results on the fourth
decimal. Using the notation CD-WAC(\(N_{\text{core}},N_{\text{valence}}, 200\)), we recommend CD-WAC(20, 20, 200)
as the default settings of the method. We remind the reader here that the order of the Padé approximant is determined by the sum \(M=N_{\text{core}}+N_{\text{valence}}+N_{\text{Im}}\). The validity of this choice will be reinforced in the next section, when discussing the screened Coulomb interaction and self-energy matrix elements.
 
\begin{figure*}[t!]
  \centering
  \begin{minipage}[b]{.4\textwidth}
    \includegraphics[scale=.33]{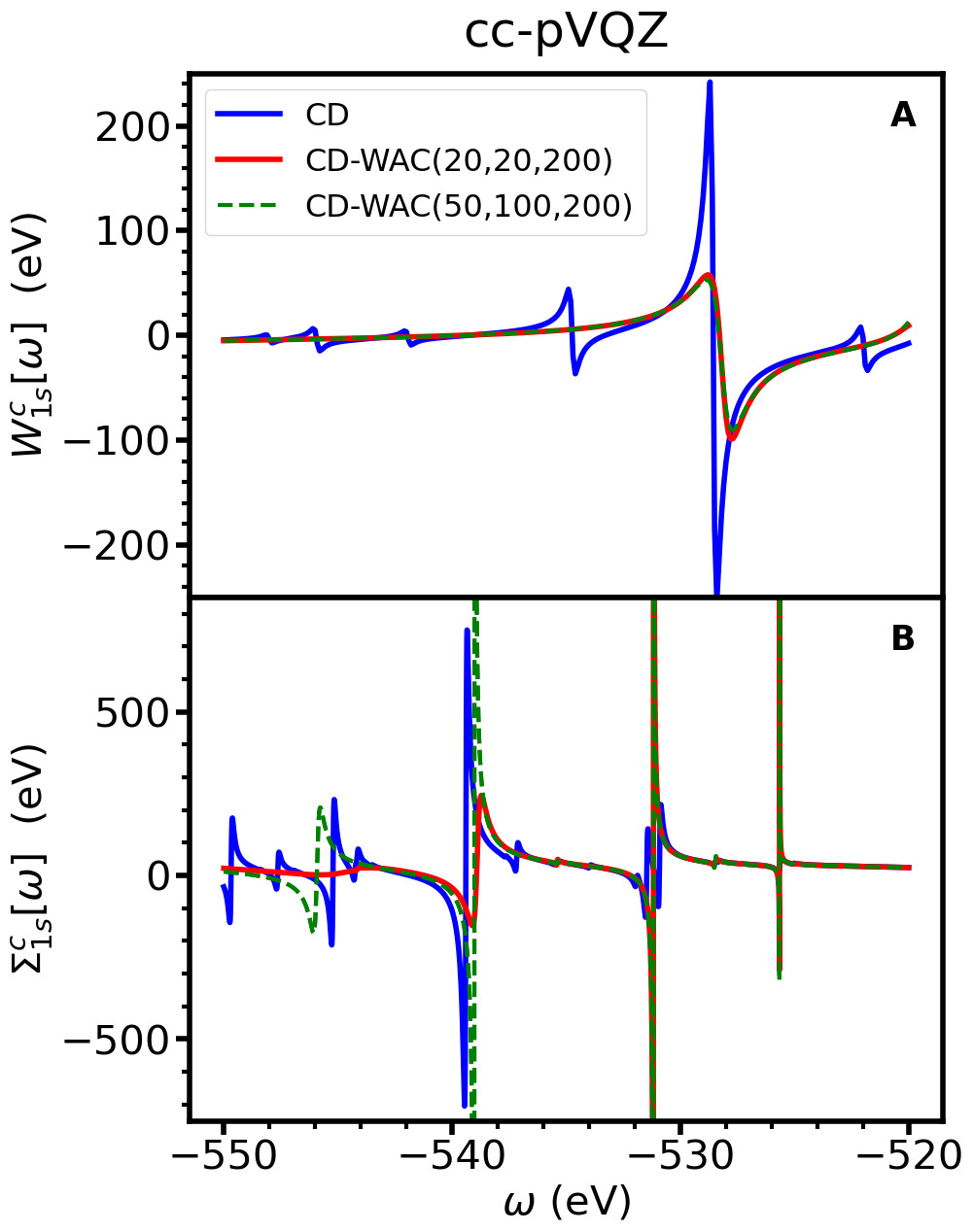}
  \end{minipage}\qquad\qquad
  \begin{minipage}[b]{.4\textwidth}
    \includegraphics[scale=.33]{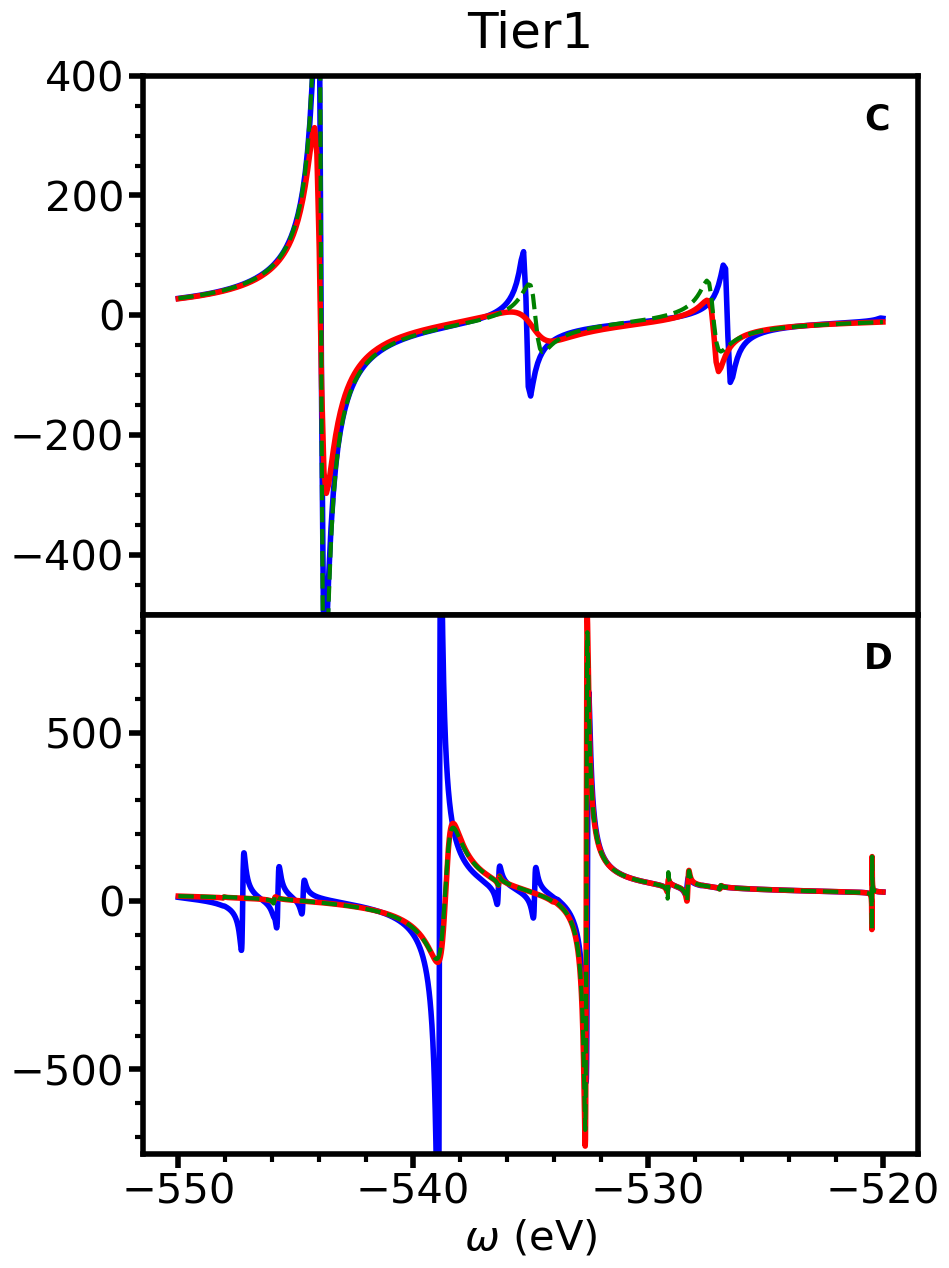}
  \end{minipage}
  \caption{Correlation part of the screened Coulomb interaction matrices \(W_{nm}^c(\omega)\) (A, C) and
    self-energy matrices \(\Sigma_{n}^c(\omega)\) (B, D) as a function of the frequency
    for the oxygen 1s core level of the
    \ce{H2O} molecule. The quantities are computed
    at the $G_0W_0$@PBE level using the cc--pVQZ (left panel) and Tier1 basis sets (right panel).}\label{fig:w_se_h2o}
\end{figure*}
 
\subsection{Screened Coulomb and self-energy structures}\label{sec:core}
 
An accurate prediction of the QP energies requires that the CD-WAC method correctly reproduces the matrix elements of the screened Coulomb and ultimately the self-energy. We focus on deep core level
matrix elements because i) \(\Sigma_n^c(\omega)\) and 
\(W_{nm}^c(\omega)\) have a less complicated pole structure for valence states\cite{golze2018core} and are easier to reproduce than their core-level counterpart, and ii) the target excitations of this work are deep core states. To facilitate comparison with relevant previous
works~\cite{golze2018core,duchemin2020robust}, we compute the
screened Coulomb matrix and the self-energy for the oxygen 1s core state of the \ce{H2O} molecule. In addition, we also inspect the influence of the basis set on the obtained curves.
 
Figure~\ref{fig:w_se_h2o} represents the real part of the \(\Sigma_n^c(\omega)\) and \(W_{nm}^c(\omega)\) matrices for \(n,m=\text{O}1s\). We compare our default settings CD-WAC(20, 20, 200) and larger grids CD-WAC(50, 100, 200) with the CD results in every case. The \(G_0W_0\) calculations on the left panel were carried out using a Gaussian basis set (cc--pVQZ). In the right panel, we used NAOs of the first tier of quality. The underlying DFT functional is PBE in all cases. PBE is not used as starting point for core-level calculations with $G_0W_0$ as discussed in Section~\ref{sec:intro}. However, among the different $GW$ flavors discussed here, the $G_0W_0@$PBE self-energies have typically the most complicated structure in the frequency region where the QP solution is expected. That means if we can reproduce the $G_0W_0@$PBE self-energy than, e.g., the one from $G_0W_0@$PBEh($\alpha$=0.45) is even more likely to be reproduced well.
 
The first noticeable difference between CD and CD-WAC is the amplitude
of the poles, which differs in both \(\Sigma_n^c\) and
\(W_n^c\). However, this difference is only a numerical artifact
introduced by the size of the imaginary component in
Equation~\eqref{eq:piriv}. Indeed, for CD calculations we only have the
infinitesimal \(i\eta\), but for CD-WAC we also have the contribution
of the imaginary part of the additional reference
frequencies.
 
More important, we can see that CD-WAC provides a highly accurate
position of the poles of both \(\Sigma_n^c(\omega)\) and
\(W_{nm}^c(\omega)\) with respect to CD, irrespective of the nature
of the basis set employed. Some smaller poles are not retrieved though, but considering that the QP energies on this example have absolute errors of \(~10^{-4}\) eV, we can conclude that either these features are simply not relevant for our target accuracy or a
product of numerical noise introduced by the discontinuous virtual-state spectrum of localized basis sets. These numerical effects are likely related to charge neutral valence excitations to high energy virtual states, leading to a few additional small poles in the core region. This is an artifact of finite basis sets, and the spectrum in the unoccupied states should become continuous in the infinite basis set limit. In a way, CD-WAC acts like a pruning device that eliminates spurious pole features in \(\Sigma_n^c(\omega)\) and \(W_{nm}^c(\omega)\) while keeping the desired accuracy, much like a systematic multipole ``plasmon pole'' approximation. We can also observe that there is little difference in the form of the curves obtained with CD-WAC(20, 20, 200) and those of  CD-WAC(50, 100, 200), which further validates our sensible defaults selection.
 
\begin{figure}[t]
  \centering
  \includegraphics[scale=.11]{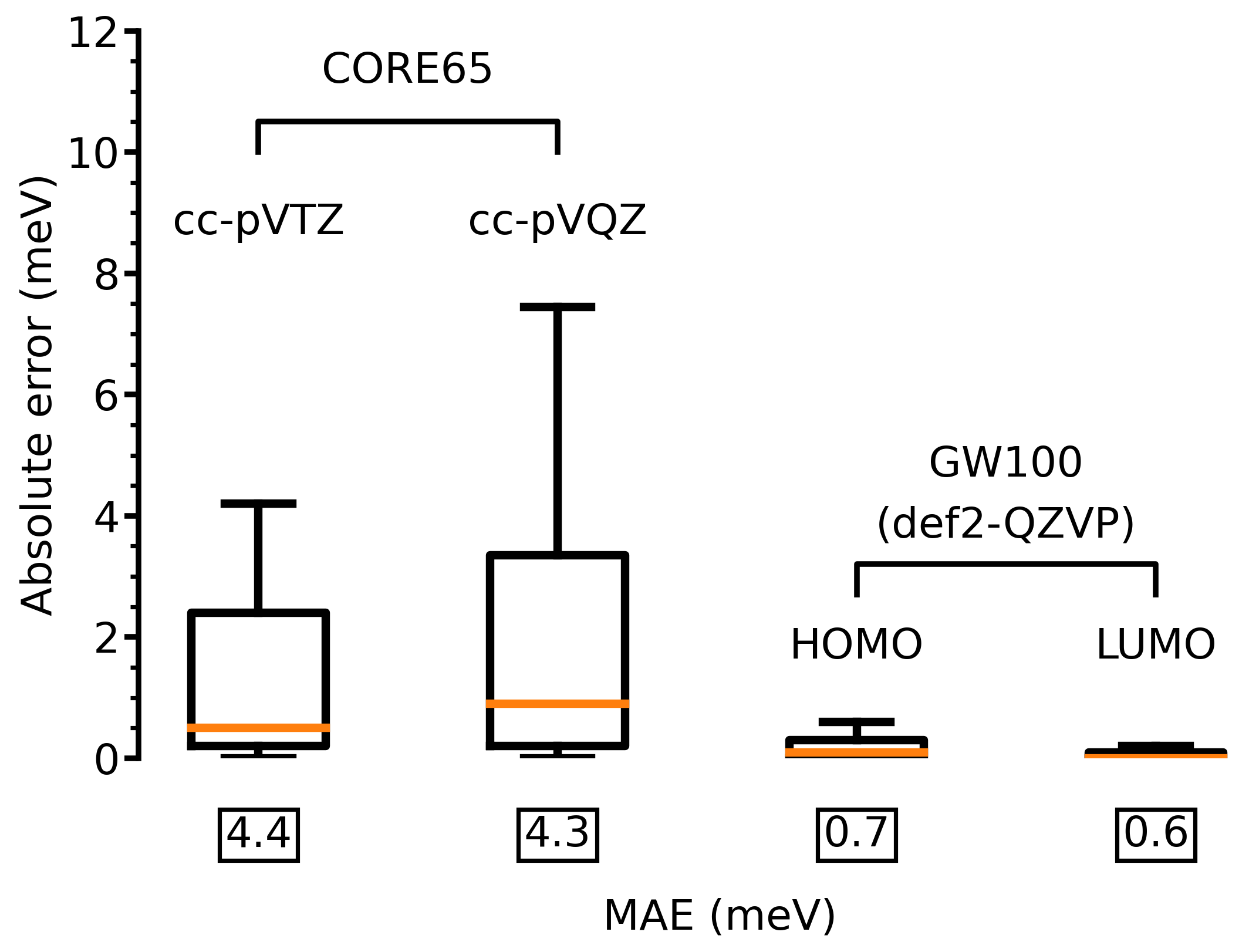}
  \caption{Box plots of the absolute errors of the CD-WAC method with respect to CD for the CORE65 and GW100 benchmark sets at the $G_0W_0@$PBEh($\alpha$=0.45) and $G_0W_0@$PBE level, respectively. The corresponding MAEs and basis sets are also reported. Boxes indicate the interquartile range, measuring where the bulk of the data are. An orange line indicates the median of the distributions.\label{fig:boxp}}
\end{figure}
 
\begin{figure}[t]
  \centering
  \includegraphics[scale=.36]{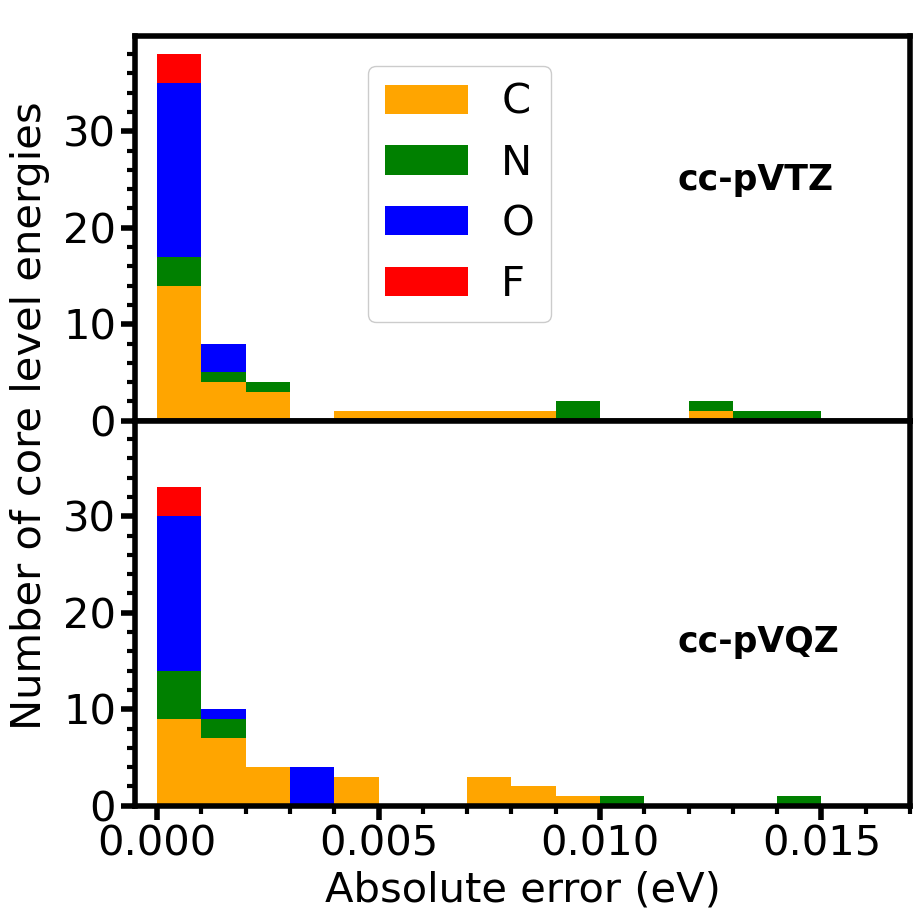}
  \caption{Histogram of the absolute errors between CD and CD-WAC per atom type in the
    CORE65 test set at the $G_0W_0@$PBEh($\alpha$=0.45) level using the cc--pVTZ and cc--pVQZ basis
    sets. Outliers are not included in the plot.\label{fig:histo}}
\end{figure}
\subsection{GW100 and CORE65 benchmarks}\label{sec:gw100}
 
In the following, we benchmark the accuracy of the CD-WAC method employing the popular GW100~\cite{van2015gw} set for frontier orbitals 
and the CORE65~\cite{golze2020accurate} set for 1s excitations. In addition to the MAEs already given in Section~\ref{sec:calib}, we analyze their dependence on the different basis sets and excitation types and report the spread of errors. 
 
Figure~\ref{fig:boxp} represent the box plots of the absolute errors between CD and CD-WAC for both GW100 and CORE65. For GW100 we use $G_0W_0@$PBE and the def2-QZVP basis set, enabling comparison with the reference data established by Setten \textit{et al.}~\cite{van2015gw}, and CD-WAC(0,0,200) settings. The HOMO and LUMO energies of the GW100 set have MAEs of the order of \(10^{-4}\) eV, and they have a symmetric, narrow distribution. The very small dispersion is confirmed by the median absolute deviation (MADs) of \(10^{-4}\) eV, which coincides with the MAEs. We also note that our CD and CD-WAC GW100 results deviate on average less than 5~meV from the data in the original GW100 paper\cite{van2015gw}, i.e., FHI-aims with analytic continuation using 16 Padé parameters and the Turbomole no-RI results.
 
All the CORE65 calculations were carried out at the $G_0W_0@$PBEh~($\alpha$=0.45) level of theory, using our CD-WAC(20,20,200) default settings. We employ the cc--pVTZ and cc--pVQZ basis sets to explore the influence of increasing basis set size on the approximation. CD-WAC yields with an MAE of 4~meV an excellent accuracy, although the error is one order of magnitude higher
than in the valence case. However, the MAEs are nicely preserved when increasing the basis set size. Compared to GW100, the distributions are wider and more skewed, particularly for cc--pVQZ. The medians are for both basis sets very close to the first quartile and the minimum. This implies that the absolute errors are below 4~meV for most 1s excitations, which is more than sufficient for the prediction of core-level excitations.
 
To further study the distribution of the absolute errors in the
CORE65 calculations, we have complemented the box plots with the histograms in
Figure~\ref{fig:histo}. These plots indeed confirm that most systems have errors below 4~meV. The histograms additionally show the 
dependence on the core-level type. We find that the N1s excitations have more frequently higher errors, while CD-WAC reproduces the O1s states best. This is also supported by Table~\ref{tab:maes_and_mads}, where the MAEs and MADs are given by excitation type. The subgroup of N1s excitation has for both basis sets MAEs of $~\sim$10~meV, which is higher than the overall MAE, but still excellent. 
 
Outliers with errors larger 0.017~eV are not displayed in Figure~\ref{fig:histo}. For cc--pVTZ, there are four outliers: three with an error of 0.02~eV and one with 0.07~eV. For cc--pVQZ, we find three outliers with errors between 0.02~eV and 0.06~eV; see SI for details. The smallest chemical shifts for second-row elements, in particular for C1s, are 0.1~eV. We note here that even the largest outliers are below this threshold.
 
\begin{table*}[t]
 \fontsize{10}{12}\selectfont
\begin{tabular}{ccccccccccc}
\toprule
\multicolumn{1}{c}{} & \multicolumn{2}{c}{cc--pVTZ} & \multicolumn{2}{c}{cc--pVQZ} &  \multicolumn{6}{c}{cc--pVTZ} \\ \midrule
\multicolumn{1}{c}{} & \multicolumn{2}{c}{$G_0W_0@$PBEh($\alpha$=0.45)} & \multicolumn{2}{c}{$G_0W_0@$PBEh($\alpha$=0.45)} & \multicolumn{2}{c}{e$vGW_0@$PBE} & \multicolumn{2}{c}{ev$GW@$PBE} & \multicolumn{2}{c}{$G_{\Delta\mathrm{ H}}W_0@$PBE} \\ \midrule
  & MAE   & MAD & MAE & MAD & MAE & MAD & MAE & MAD & MAE & MAD \\
all & 4.408 & 0.4000 & 4.315 & 0.8000 & 147.7 & 48.92 & 36.81 & 6.450 & 26.25 & 0.4000  \\
C &  4.700 & 1.100  & 3.571 & 1.400 & 71.21 & 31.42 & 54.48 & 7.210 & 11.20 &  1.600 \\
N & 12.41 & 6.800 & 14.01 & 1.150 & 101.7 & 56.50 & 32.55 & 4.450 & 91.37 &  19.55 \\
O & 0.4071 & 0.1500 & 0.9024 & 0.3000 &  257.5 & 57.85 & 17.36 & 5.300 & 17.35 &  0.1000 \\
F & 0.1417 & 0.02500 & 0.06667 & 0.0000 & 312.9 & 34.97 & 11.76 & 7.475 & 0.2722 &  0.05000 \\\bottomrule
\end{tabular}
\caption{MAEs and MADs in MeV for CD-WAC with respect to CD using the CORE65 benchmark set and different \(GW\) flavors.}\label{tab:maes_and_mads}
\end{table*}

\subsection{Beyond \boldsymbol{$G_0W_0$} calculations}
 
Next, we assess the quality of the CD-WAC approximation for methods
beyond \(G_0W_0\). We performed ev\(GW_0\), ev\(GW\),
and \(G_{\Delta \text{H}}W_0\) calculations using CD-WAC(20, 20, 200) and
compared them with CD. Table~\ref{tab:maes_and_mads} contains the MAEs and MADs for each excitation type, as well as the overall values. We remind the reader that the statistics for F1s are rather poor since we have only three of those excitations in the benchmark set, but we included it for the sake of completeness.

Compared to \(G_0W_0\), the partially self-consistent schemes have larger MAEs. For  ev\(GW\) and \(G_{\Delta \text{H}}W_0\) we observe an increase by an order of magnitude and even two orders of magnitude for  ev\(GW_0\).   
A detailed report of the individual core excitations is provided in the Tables S2 and S3 of the Supplementary Information (SI).
 
Starting the discussion with ev\(GW\), we find that the different excitation types have similar MAEs, each $<$55~meV. In comparison to \(G_0W_0\), the MADs are consistently larger by a factor of 10, indicating that also the dispersion of the data increases. The reason for the increase might be attributed to the fact that the PBE self-energies tend to have more features than the ones from a hybrid functional such as PBEh, making it more difficult to reproduce them by CD-WAC. 
 
While the accuracy of CD-WAC is still satisfying for ev\(GW\), it is surprisingly large for ev\(GW_0\). With an MAE of 0.1~eV, the CD-WAC error touches the experimental resolution of carbon 1s binding energies. The  lower errors observed in ev\(GW\) offer a reason for this impairment. The \(W_{nm}^c\) matrices are recomputed in every iteration of ev\(GW\) using the perturbed eigenvalues, including the computation of the extra references. This results in a new Thiele interpolant being constructed in each iteration, which compensates for the error introduced in the computation of the updated eigenvalue. However, in ev\(GW_0\), only one interpolant is employed for all iterations, leading to an accumulation of error on the perturbed eigenvalue without compensation. Another source of error in ev\(GW_0\) is the interpolation vicinity, which for the points in \(\omega_V\) is centered on \(|\varepsilon_F-\varepsilon_m|\), where $m$ indexes a core states. However, after some iterations a more adequate center would be \(|\varepsilon_F-\varepsilon_m^{GW}|\) because the difference between KS eigenvalue and QP energy can be easily tens of eV for deep core states.\cite{golze2018core} This problem also does not occur in ev\(GW\).
 
It should be in principle possible to improve the accuracy for ev$GW_0$. For example, the Padé approximant could be re-evaluated after some iteration steps with  reference points from the $\omega_V$ region re-centered at \(|\varepsilon_F-\varepsilon_m^{GW}|\). However, in our recent benchmark study\cite{li2022benchmark}, we proposed the \(G_{\Delta \text{H}}W_0\) as an approximation to ev\(GW_0\), which yields the same errors with respect to experiment, but is in terms of computational cost comparable to \(G_0W_0\). For large systems, we would rather use \(G_{\Delta \text{H}}W_0\) or $G_0W_0@$PBEh($\alpha$=0.45). For small systems, the CD is computationally affordable, which leaves little incentive to tune the CD-WAC performance for ev\(GW_0\). 
 
The CD-WAC performance for \(G_{\Delta \text{H}}W_0\) is similar to ev\(GW\), yielding MAEs \(<\)100~meV. The reason for the decreased accuracy compared to \(G_0W_0\) must be again attributed to the underlying PBE functional, which yields feature-rich self-energy structures. Unlike for ev\(GW\), the MAEs differ between the atom types. In particular, the MAE for the N1s stands out, caused by four outliers with errors between $0.1 - 0.4$~eV. 
Closer inspection of the outliers  reveals the CD self-energy has shallow, small poles close to the QP solution, see for example Figure S2. 
 
As already mentioned in Section~\ref{sec:core}, we believe that these poles are artifacts caused by the discontinuous virtual-state spectrum in localized basis functions. They can be associated to charge neutral excitation $\Omega_s$ from the valence states to the highest virtual states, which occur in the self-energy, see Equation~\eqref{eq:sigma_pole}. There are several observations that support this interpretation: i) $\Omega_s$ are close to eigenvalue differences. For the outliers, the highest virtual states are indeed in the range of the N1s binding energies with cc-pVTZ. This also explains why the other excitation types are not affected.  ii) We observe these spurious poles more frequently with an underlying PBE functional than with PBEh. $\Omega_s$ are underestimated at the PBE, but overestimated at the PBEh level, shifting them out of the frequency range of the QP solution. iii) The number of these spurious poles increases with the basis set size because, e.g., the cc-pV5Z and cc-pVQ6Z basis sets span virtual states up to several thousand eV, see also the SI of Ref.~\citenum{golze2020accurate}, where we briefly discussed this issue as cause for the disturbance in the basis set extrapolation. CD-WAC prunes this spurious poles  and arguably yields the better result for localized basis functions.
 
\subsection{Performance of the implementation}
\label{sec:performance}
The key feature of the CD-WAC approach is the reduction of the computational requirements of the standard CD approach by finding an analytical model for the residue term $R^c_n$ (Equation~\eqref{eq:Rn_term}). In the valence case, we showed that it is sufficient to include only the imaginary frequency points for the analytic continuation of $W$, which implies re-using the $W_{mn}(i\omega)$ matrix elements that we compute anyway for the $I^c_n$ term (Equation~\eqref{eq:In_term}). For the HOMO, one typically has to compute one or two residues independent of the system size. Thus, CD-WAC saves the computation of $W_{mn}$ for two (real) frequency points. This corresponds to a marginal prefactor reduction compared to CD: Assuming we have an imaginary frequency grid of 100 or 200 points and that the system is large enough for the $W_{mn}$ evaluation to dominate the computational cost, then CD-WAC saves at best 1\% of the total computational time. 
 
\begin{figure}[t]
  \centering
  \includegraphics[scale=.27]{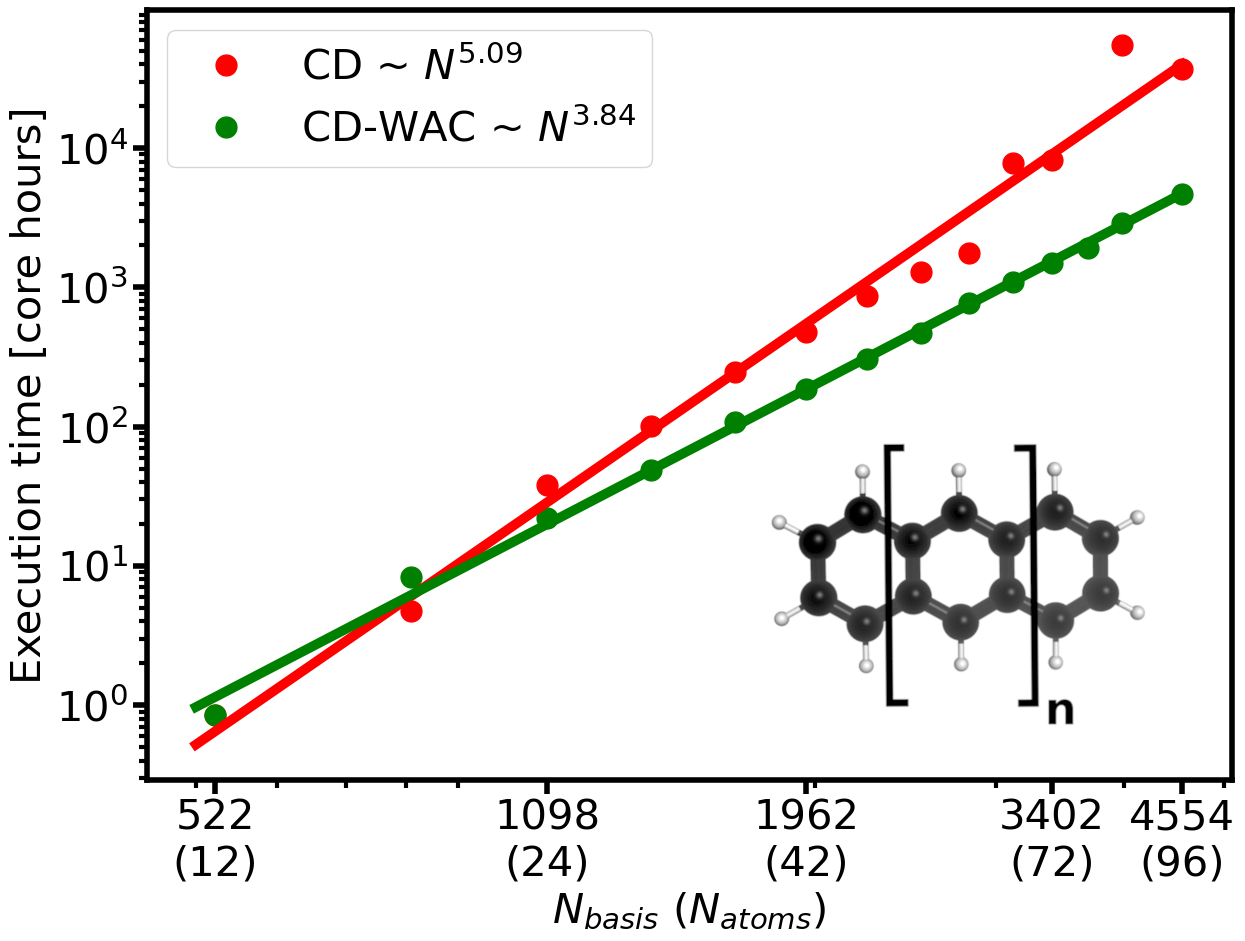}
  \caption{Performance of the CD-WAC implementation compared with CD\@. The plots report the execution time of the $GW$ part for acene chains from 2-14,15 rings using the def2-QZVP basis set. The solid lines represent the two-parameter least-squares fits of prefactor and exponent. The latter is reported in the legend of the plot. Calculations performed with Intel Xeon Platinum 8168 CPUs at 2.7 GHz.\label{fig:sca}}
\end{figure}

The situation is completely different for core states, for which the CD-WAC methods reduces the scaling by one order of magnitude from $O(N^5)$ to $O(N^4)$. We demonstrate the latter by conducting $G_0W_0@$PBEh($\alpha$=0.45) calculations for a series of acene chains \ce{(C_{4n+2}H_{2n+4})_{n=1,\ldots, 13, 15}} using CD--WAC(100, 0, 100), i.e.,  generic settings. Figure~\ref{fig:sca} displays the execution time for the $GW$ calculation, excluding the preceding DFT calculation and the cubic scaling computation of the RI integrals (Equation~\eqref{eq:3c_OnmP}). The complexities of both the CD and CD-WAC calculations precisely agree with the theoretical predictions and match the results of previous studies in the case of CD~\cite{golze2018core}. 
 
To understand the origin of the scaling laws, we need to consider the \(I_n^c\) and \(R_n^{c}\) terms independently. In both CD and CD-WAC, the numerical quadrature of the integral \(I_n^c\) is performed by using, e.g., a modified Gauss-Legendre grid with \(N_{\omega}\) points. The grid size is practically independent of the system size~\cite{ren2012resolution}.
The scaling of the quadrature will be dominated by the computation of the matrix elements \(\Pi_{PQ}(i\omega)\). Formally, this requires \(N_{\omega}N_{\text{occ}}N_{\text{virt}}N_{\text{aux}}^2\) operations, where \(N_{\text{occ}}\) and \(N_{\text{virt}}\) are the number of occupied and virtual states, respectively, and \(N_{\text{aux}}\) is the number of auxiliary (RI) basis functions. This leads to a \(O(N^4)\) complexity for the \(I_n^c\) term because \(N_{\text{occ}}\), \(N_{\text{virt}}\) and \(N_{\text{aux}}\) increase linearly with system size.  
 
The computation of the \(R_n^{c}\) term is dominated by \(\Pi_{PQ}(|\epsilon_m-\omega|)\) as previously shown in Ref.\citenum{golze2018core}. In the CD implementation, this requires \(N_{\text{res}}N_{\text{occ}}N_{\text{virt}}N_{\text{aux}}^2\)
operations, where \(N_{\text{res}}\) are the number of residues
entering the summation in equation~\eqref{eq:Rn_term}. For valence states, this number is usually small and independent of the system size, as discussed above, and the overall scaling is $O(N^4)$. For deep core states, however, it is of the order \(N_{\text{res}}\sim N_{\text{occ}}\), which gives rise to the unfavorable scaling of \(O(N^5)\). This can be  understood by inspecting the arguments of \(\Pi_{PQ}\) in \(R_n^{c}\) and \(I_n^{c}\) in equations~\eqref{eq:Rn_term} and ~\eqref{eq:In_term}. For \(R_n^{c}\), the argument of \(\Pi_{PQ}\) depends on the index \(m\), which runs over the number of residues, whereas we have a dependency on an imaginary grid of constant size \(N_{\omega}\) in \(I_n^{c}\). When the analytic continuation of \(W\) is carried out for \(R_n^{c}\), the computational time for evaluating the latter is negligible during the iteration of the QP equation~\eqref{eq:qp_eq}. The computational cost in CD-WAC is hence dominated by the \(I_n^c\) term and the quartic scaling computation of $W_{mn}^c$ for the additional real frequencies \(\bm{\mathcal{F}}_{\mathrm{real}}\). Since the number of extra points is independent on the system size, the total scaling of CD-WAC is consequently \(O(N^4)\).
 
The calculation of $W_{mn}^c$ for the additional real frequencies adds to the prefactor in the CD-WAC algorithm. For small systems, the number of residues (times the number of iteration steps of equation~\eqref{eq:qp_eq}) is smaller than the additional WAC frequencies. In this case, CD is faster than CD-WAC. However, the cross-over between CD and CD-WAC occurs already around 16 atoms as shown in Figure~\ref{fig:sca}, where generic CD-WAC settings with 100 extra points are used. With our optimized settings, only 40 additional points are required, which moves the cross-over point between CD and CD-WAC to even smaller systems. 
 
Next, we investigate the performance improvement for the total run times, including now also the DFT part and the calculation of the 3-center RI quantities \(O^{nm}_P\) (Equation~\eqref{eq:3c_OnmP}). The 3-center quantities are computed before the SCF cycle starts. They enter the computation of the  exchange energy in hybrid DFT and are then also used in the polarizability and $GW$ self-energy. As detailed in our previous work~\cite{golze2018core}, the RI integral evaluation has a cubic time complexity, but it has potentially a large prefactor dependent on the type of basis function treated in our NAO scheme. The 3-center RI integral can significantly contribute to the total time, while the overhead due to other computation steps in the DFT part is marginal. In the following, we discuss therefore only the RI and \(GW\) self-energy evaluation as relevant contributions.  
 
\begin{figure}[t]
  \hspace{-.9cm}
  \includegraphics[scale=.36]{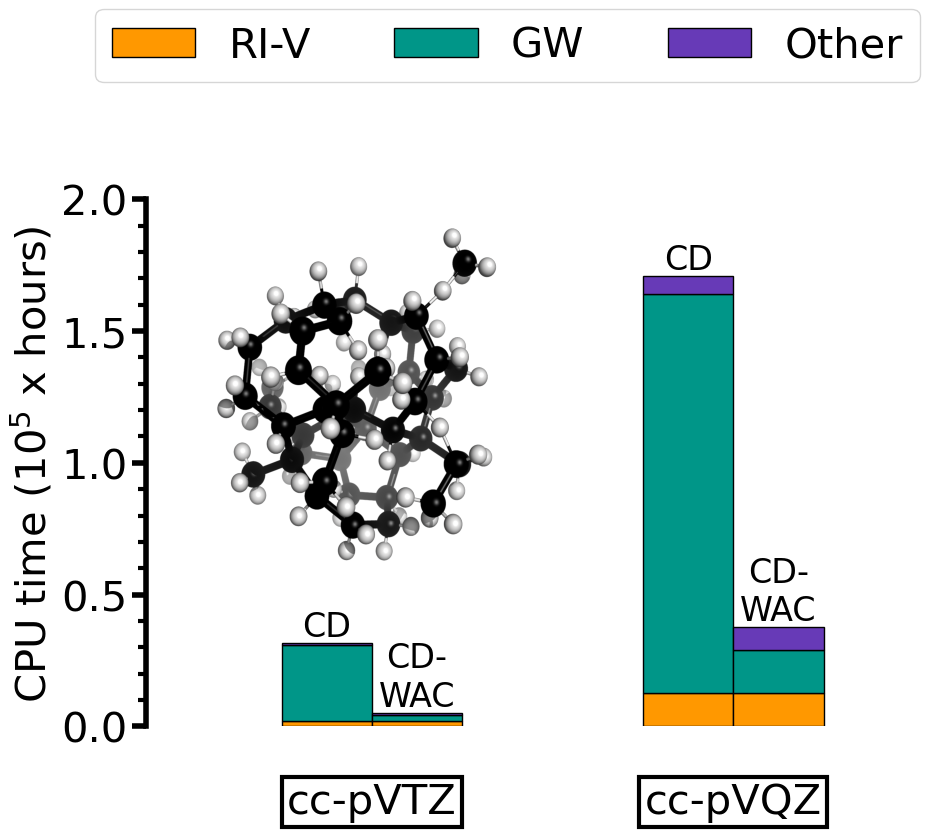}
  \caption{
    Comparison of the total run time (CPU hours) for the computation of the C1s excitation of the central atom in the amorphous carbon cluster \ce{a-C46H70} using CD and CD-WAC(20, 20, 200). The absolute errors (CD vs. CD-WAC) for the cc--pVTZ and cc--pVQZ are 0.3~meV and 0.4~meV, respectively. Calculations performed with Intel Xeon Platinum 8168 CPUs at 2.7 GHz.\label{fig:amorph}}
\end{figure}
 
We investigate the CD-WAC speed-up for the total execution time for a production-run system, namely an amorphous carbon cluster of 116 atoms (\ce{a-C46H70}). The \ce{a-C46H70} cluster is part of the $GW$ training data, which we used to develop a machine-learning model for XPS predictions of disordered carbon materials.\cite{golze2022accurate} Figure~\ref{fig:amorph} shows the total execution time for the spin-polarized calculation of the C1s excitation of the central carbon atom in the cluster. We compare CD and CD-WAC(20, 20, 200) and indicate with different colors the RI and $GW$ contributions to the run time. 
The settings are the same as in our previous work\cite{golze2022accurate}, i.e. Dunning basis sets and $G_0W_0@$PBEh\((\alpha=0.45\)). 
CD-WAC speeds-up the $GW$ self-energy evaluation by a factor of 12.6 and 9.2 for cc-pVTZ and cc-pVQZ, respectively. The RI part, which takes up 6\% (cc-pVTZ) and 12\% (cc-pVQZ) of the total CD run time, is unaffected by the improvements of CD-WAC. For the total run, we attain a speed-up by a factor of 6.0 and 4.5. The difference in the speed-up between cc-pVTZ and cc-pVQZ is likely introduced by memory bottlenecks created by the inter-process communications involved in the computation and usage of the RI-V tensors $O^{nm}_P$. An indication for the latter is that the contribution of RI to the total CD run time is larger for cc-pVQZ than for cc-pVTZ. Nevertheless, the speed-ups are substantial. It should be emphasized that CD-WAC retains the high accuracy for larger systems with errors $< 1$~m~eV. 
 
An additional benefit of the CD-WAC method is its ability to expedite the computation of the spectral function. The spectral function $A(\omega)$ is derived from the imaginary part of the Green's function, and its diagonal elements are given by $\displaystyle{A_{nn}(\omega) ~\propto \text{Im}\left[(\omega - \epsilon_n-\Sigma_n(\omega)+v^{xc}_n)^{-1}\right]}$\cite{golze2019gw}. The spectral function 
provides the full spectral information, including access to the satellite spectrum. However, the computation of $A(\omega)$ is burdensome with CD since it requires an evaluation of $\Sigma_n^c$ for each frequency point $\omega$. If we have a frequency range of 50~eV and intervals of 0.01~eV, we compute the self-energy 5k times. In CD-WAC, we have an analytic approximation to the $R_n$ term and only need to  perform the numerical integration in Equation~\eqref{eq:In_term} with pre-computed matrix elements $W^c_{nm}(i\omega)$. Even for a small system, such as a single water molecule, CD-WAC reduces the computational cost for the computation of $A(\omega)$ by at least three orders of magnitude compared to CD\@.

The advantageous performance of CD-WAC for computing spectral functions and self-energy matrix elements also becomes apparent when compared with similar algorithms that successfully reduce the scaling of the original CD\@. As an example, the CD-MINRES method~\cite{mejia2021scalable} can accurately compute the binding energies for the CORE65 benchmark set, while having a formal \(O(N^4)\) time complexity. However, the evaluation of the \(R_n\) term of CD-MINRES still requires performing several matrix computations, whereas in CD-WAC one only needs to evaluate the analytic expression of the continued fraction.
 
\section{Conclusions}\label{sec:conc}
 
We have presented an efficient and scalable \(GW\) implementation for core-level calculations. Building on our previous work\cite{golze2018core}, we use the highly accurate and numerically stable CD method to compute the \(GW\) self-energy for the full-frequency range on the real axis. In the present work, we have addressed the computational bottleneck of CD for deep core excitations by combining it with the AC of the \(W\) matrices, which is carried out using a modified version of Thiele's reciprocal differences' algorithm. 
 
We have found that it is more difficult to enable the CD-WAC method for core than valence levels: i) We have devised an algorithm, which numerically stabilizes the Pad\'{e} approximation for binding energies $>$ 100~eV. ii) While a set of imaginary frequency points is sufficient for valence states, real frequency points must be additionally included in the AC of $W$ when treating inner-shell excitations. We have implemented a heuristic procedure which places the additional real frequency in two different frequency regions, which are defined by the core and valence residues. Using the notation CD-WAC(\(N_{\text{core}},N_{\text{valence}},N_{\text{Im}} \)), we recommend CD-WAC(20,20,200) as safe default settings.
 
We have comprehensively benchmarked the CD-WAC approach against CD for 1s excitations. We have demonstrated that CD-WAC reproduces the essential features of the $W$ and $\Sigma$ matrix elements regardless of the type of the local basis set. We have studied the CORE65 benchmark set with $G_0W_0@$PBEh and partially self-consistent schemes, namely ev$GW_0$, ev$GW$ and $G_{\Delta\text{H}}W_0$, using PBE as starting point. For $G_0W_0@$PBEh, CD-WAC yields MAEs of 4~meV with respect to CD, independent on the basis set size. The error increases for the self-consistent schemes, but is with MAEs $< 50$~meV still excellent for ev$GW$ and $G_{\Delta\text{H}}W_0$. In the ev$GW_0$ case, the MAE is with 0.1~eV already in the range of the chemical C1s shifts. The reduction of the error is in principle possible. However, the CD-WAC approach is designed for large-scale  calculations, for which we devised $G_0W_0@$PBEh\cite{golze2020accurate} and more recently $G_{\Delta\text{H}}W_0$\cite{li2022benchmark} methods as computationally affordable alternatives to ev$GW_0$. 
 
The computational performance of CD-WAC has been demonstrated through numerical experiments. Compared to CD, the CD-WAC approach reduces the scaling for core-level $GW$ calculations from $O(N^5)$ to $O(N^4)$ with respect to system size $N$. The prefactor introduced by computing the additional reference matrices for the AC of $W$ is small, and the cross-over point between CD and CD-WAC is around 20 atoms. The speed-up has been assessed for a medium-size system of 116 atoms, for which we found that CD-WAC accelerates the total run time by at least a factor of 5 to 6.  
 
The CD-WAC method represents a valuable contribution to the current efforts of reducing the scaling of \(GW\) calculations and paves the way for accurate computational predictions of core-level excitations in complex condensed matter systems. Further scaling reduction and the extension to periodic systems are part of ongoing and future work. Moreover, the CD-WAC dramatically speeds up the computation of the spectral function. The latter gives access to the satellite spectrum, which provides valuable complementary information to QP excitations.
 

\begin{acknowledgement}
\fontsize{10}{12}\selectfont
  The authors acknowledge financial support by the Emmy Noether Programme of the German Research Foundation (project number 453275048) and thank the ZIH of the TU Dresden,
  the Jülich Supercomputer Computer Center and the Finnish CSC - IT Center for Science for providing computational resources.
\end{acknowledgement}
 
\begin{suppinfo}
 
 The supplementary information is available free of charge. We include the tables comparing the CD and CD-WAC methods for different \(GW\) flavors (Tables S1 to S3). We also present the box plots of the CD-WAC absolute errors with respect to CD at the \(G_0W_0\)PBEh($\alpha$=0.45) level of theory for increasing sizes of additional frequency points (Figure S1), and the self-energy matrix elements for the pyrrole molecule for both CD and CD-WAC at the $G_{\Delta H}W_{0}@$PBE level of theory (Figure S2).
 
\end{suppinfo}
 
\bibliography{references}
 
\end{document}